# Unsupervised Segmentation-Based Machine Learning as an Advanced Analysis Tool for Single Molecule Break Junction Data

Nathan D. Bamberger[†], Jeffrey A. Ivie[†], Keshaba N. Parida[†], Dominic V. McGrath[†], and Oliver L.A. Monti[†,§,*]

[†]Department of Chemistry and Biochemistry, University of Arizona, 1306 E. University Blvd., Tucson, Arizona 85721, USA

[§]Department of Physics, University of Arizona, 1118 E. Fourth Street, Tucson, Arizona 85721, USA

ABSTRACT: Improved understanding of charge-transport in single molecules is essential for harnessing the potential of molecules *e.g.* as circuit components at the ultimate size limit. However, interpretation and analysis of the large, stochastic datasets produced by most quantum transport experiments remains an ongoing challenge to discovering much-needed structure-property relationships. Here, we introduce Segment Clustering, a novel unsupervised hypothesis generation tool for investigating single molecule break junction distance-conductance traces. In contrast to previous machine learning approaches for single molecule data, Segment Clustering identifies groupings of similar *pieces of traces* instead of *entire traces*. This offers a new and advantageous perspective into dataset structure because it facilitates the identification of meaningful local trace behaviors that may otherwise be obscured by random fluctuations over longer distance scales. We illustrate the power and broad applicability of this approach with two case studies that address common challenges encountered in single molecule studies: First, Segment Clustering is used to extract primary molecular features from a varying background to increase the precision and robustness of conductance measurements, enabling small changes in conductance in response to molecular design to be identified with confidence. Second, Segment Clustering is applied to a known data mixture to qualitatively separate distinct molecular features in a rigorous and unbiased manner. These examples demonstrate two powerful ways in which Segment Clustering can aid in the development of structure-property relationships in molecular quantum transport, an outstanding challenge in the field of molecular electronics.

## 1. INTRODUCTION

Ever since 1974, when Aviram and Ratner proposed using a single molecule to rectify current,[1] the nanoscale transport community has pursued the goal of molecular-based circuitry to take advantage of the small size, enormous design space, and potential low manufacturing costs of circuit components composed of individual molecules.[2] However, in order to create functional devices that can capitalize on these advantages, it is first necessary to understand the fundamental physics and design principles underlying charge transport in single molecule systems. This understanding is most commonly gained using either Mechanically Controlled Break Junctions (MCBJs)[3–10] or Scanning Tunneling Microscope Break Junctions (STM-BJs),[11–15] techniques which pull apart a thin metal bridge, typically made from gold, to form a nanogap, while simultaneously applying a small bias across the bridge or gap and recording the resulting current. The changes in current when individual molecules bridge the gap provide insight into the electrical non-equilibrium properties of single-molecule circuit components.

Most commonly, such experiments yield "breaking traces", in which the junction conductance $G = I/V$ is recorded as a function of stretching distance during the breaking process. Figure 1a contains example breaking traces collected using our MCBJ set-up with the molecule OPV3-2BT-H (Chart 1), plotted on a log-linear scale in order to capture the large dynamic range of possible molecular conductances, as is standard in the field. These examples illustrate three characteristic features of breaking traces: 1) Just before rupture, a plateau occurs at the conductance value corresponding to a single atomic point contact. For Au electrodes, this value is 77.48 μS,[16] and denoted 1 $G_0$; 2) When no molecule is bound in the junction (blue traces), the conductance is solely due to tunneling and decays exponentially; and 3) When a molecule *is* bound in the junction (red traces), the conductance is roughly constant (though potentially fluctuating) over the length of the molecule, forming a "molecular plateau".

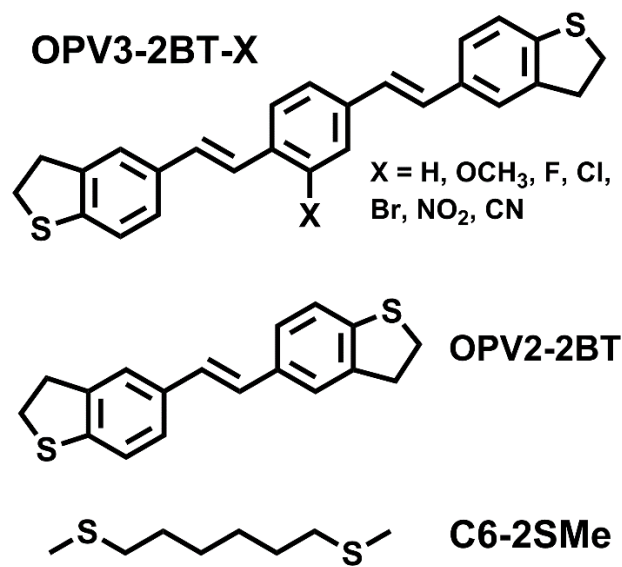

Chart 1. Structures of molecules considered in this work and their associated abbreviations.

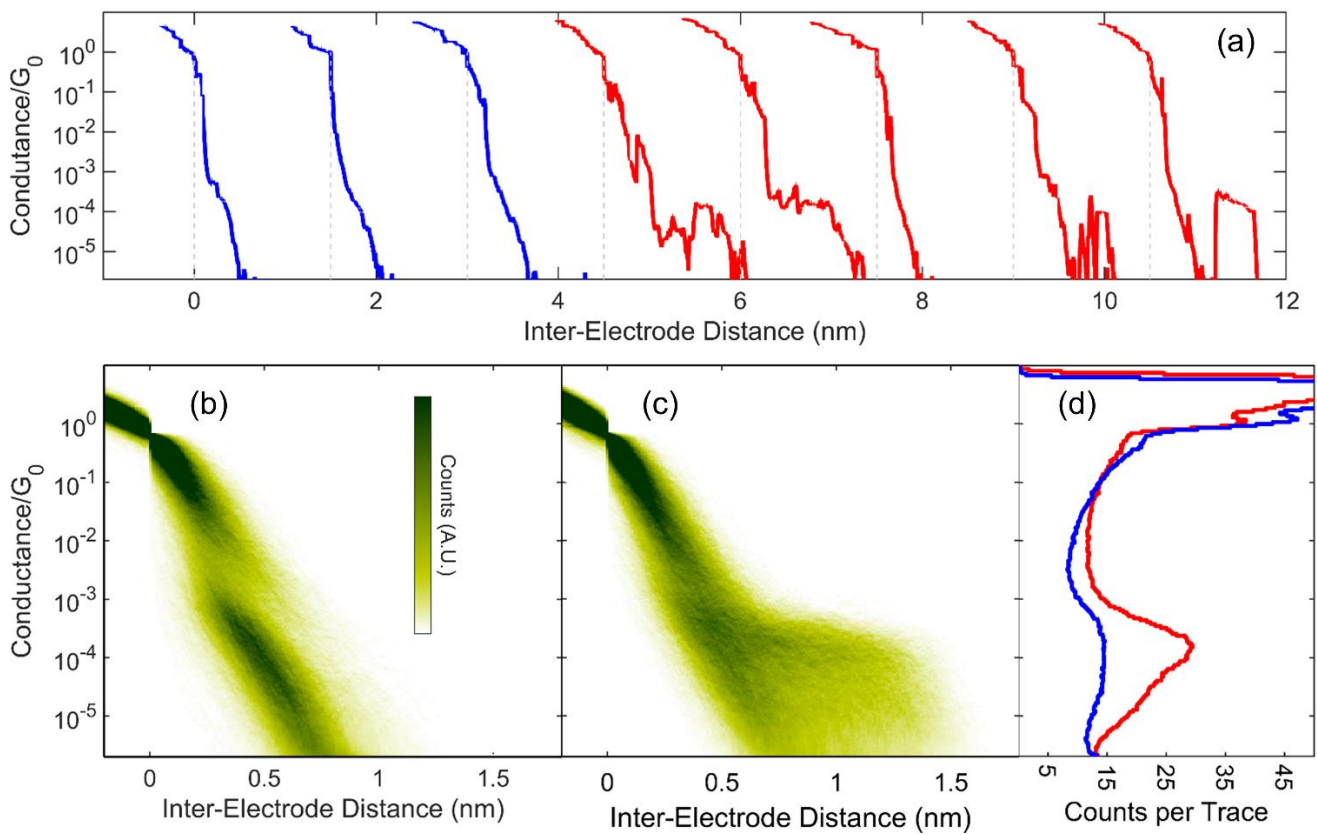

**Figure 1.** Break junction data collected with the molecule OPV3-2BT-H. (a) Selected breaking traces from before (blue) and after (red) the addition of molecules, offset by 1.5 nm for clarity. The blue traces illustrate exponential tunneling decay in an empty nanogap (linear on a logarithmic scale), while the red traces illustrate molecular plateaus and their variability. (b,c) 2D histograms of 7122 and 6280 consecutive breaking traces collected before and after the addition of molecules, respectively. (b) exhibits a clear tunneling decay feature below $10^{-3}$ $G_0$, while (c) exhibits a pronounced molecular feature extending out to ~1.5 nm at ~$10^{-4}$ $G_0$. (d) 1D histograms for the datasets in (b) (blue) and (c) (red). While both histograms display a sharp peak at ~1 $G_0$ from the single gold point contact plateaus, only the histogram collected after molecular addition displays a broad peak at ~$10^{-4}$ $G_0$ due to presence of molecules.

However, because of the stochastic nature of the breaking process, molecular conformation, and molecular diffusion in and out of the junction, individual molecular traces are highly variable. In particular, plateaus for the same molecule can vary by over an order of magnitude in conductance (*e.g.* first two red traces in Fig. 1a); some traces collected in the presence of molecules do not display any molecular plateau at all (*e.g.* third red trace in Fig. 1a); and molecular plateaus may break off and re-form within the same trace (*e.g.* last two red traces in Fig. 1a). In order to capture this variability, thousands of traces are collected under the same experimental conditions. A set of traces can then be summarized by a 2D histogram (Fig. 1b,c), which shows the frequency of observing each pair of inter-electrode distance and log(conductance) values; or a 1D histogram (Fig. 1d), which is obtained from the 2D histogram by integrating out the inter-electrode distance dimension to "collapse" all of the data onto the log(conductance) axis.

While such histograms usefully summarize the ensemble of single molecule conductance behaviors, they obscure likely meaningful differences within and among different molecular constructs that could be harnessed to advance a host of intriguing molecular electronics research directions. At present, 1D histograms are often used to determine a single "peak" or "most probable" conductance for a given molecule,[17–27] and 2D histograms have been used to separate molecular features that may correspond to distinct physical phenomena, such as different binding modes.[9,28–32] However, the broad features found in these histograms make it difficult to confidently separate features without introducing bias, and the complex "background" signature, composed of tunneling decay and broken molecular plateaus, makes it hard to robustly fit molecular peaks. These inter-related challenges have motivated several research groups to develop automated clustering and data-sorting methods for analyzing breaking traces[33–41] and related data.[15,42–44] Broadly speaking, the goal of these approaches is to partition a large dataset of highly-varied traces into separate groupings in order to improve the robustness of peak conductance measurements and/or to identify distinct junction behaviors. Using an automated algorithm to identify clusters of data helps eliminate bias towards seeing only the types of groupings that are expected *a priori*. The clustering approaches developed so far are based on techniques ranging from principal component analysis[34,38] to neural networks,[35,37,38,41,43] and have found success in separating known features in experimental or simulated data,[34,35,37,42,43] and in detecting intriguing sub-features for further quantitative or qualitative analysis.[33,34,38,41,42,45]

Nearly every published clustering approach applied to breaking traces treats each entire trace as one single object.[15,33–35,37–42,44] This choice implicitly assumes that the overall trajectories that traces follow are non-random, and hence such algorithms are best suited for traces that exhibit few unpredictable fluctuations. However, our own experimental data and many published examples suggest that this is often only true over distances much shorter than most molecular lengths. Over longer distances, there are often sudden and unpredictable conductance shifts between mostly linear sections,[32,46–50] and in some instances such traces constitute the majority of all molecular "plateaus". Whole-trace focused methods can thus easily miss a meaningful sub-feature, even one conserved across many traces, if the other parts of those traces differ significantly due to random and uncorrelated behavior. We therefore designed a new approach, "Segment Clustering", based on the idea of defining *pieces of traces* as the objects to be clustered, and in particular linearly approximated segments. This definition better matches the empirical structure of trace trajectories in most systems studied so far,[13,51–59] ranging from *in situ* chemical reactions to photo-switching. Segment Clustering is thus able to identify the truly conserved features in highly-stochastic datasets and has the potential to reveal insights not available to other clustering approaches. Additionally, Segment Clustering does not require training, like some neural network-based approaches,[35,38,41,43] nor does it rely on criteria that are likely dataset-specific, like many filtering-based approaches,[15,40,44] and so is expected to be easily generalizable to new datasets.

We emphasize that Segment Clustering is neither expected nor designed to identify *every* meaningful feature in *every* single molecule dataset. Instead, it focuses on one broad category of features—approximately linear trace sections—which are evidently quite common in distance-conductance traces, thus providing a new perspective into dataset structure. At the same time, just because segment clustering identifies a given cluster does not, by itself, constitute proof that such a cluster corresponds to a distinct physical behavior. Rather, Segment Clustering is designed as a *hypothesis generation tool*: by identifying data groupings that may not be obvious to the naked eye and which do not rely on preconceived and potentially flawed notions of meaningful data structure, it can help spawn ideas of what types of behaviors may be present in single-molecule junctions. These ideas can then be tested *via* additional experiments or targeted data analysis, laying the basis for further insight into the fundamental physics of single-molecule transport.

In the remainder of this paper, we describe our experimental methodology and then explain in detail the motivation and mechanics behind Segment Clustering. We next present two case studies using our own MCBJ data to illustrate two applications of Segment Clustering. In the first case study, we show that Segment Clustering can reliably separate the "primary" molecular feature from a shifting background signal, enabling us to confidently distinguish small changes in conductance across a family of similar molecules. In the second case study, we use a known data mixture to demonstrate that Segment Clustering can separate molecular features even when they come from overlapping conductance distributions.

2. EXPERIMENTAL SECTION

**2.1. Fabrication.** Samples for the MCBJ experiment were fabricated by depositing a gold wire on a polyimide-coated phosphor bronze substrate using electron beam evaporation. A 4 nm titanium layer was used to improve adherence of the 80 nm thick gold film. The pattern for gold deposition, including an ~100 nm wide gold bridge in the center of the wire, was fashioned by electron beam lithography. The gold bridge was then created *via* $O_2/CHF_3$ plasma etching of the polyimide to produce an ~1-2 μm undercut (Fig. S1a,b).

**2.2. Trace Collection.** Samples were clamped and then bent with a push rod placed underneath the gold bridge (Fig. S1c). A 100 mV bias was applied across the gold bridge while simultaneously measuring the conductance of the bridge using a custom, high-speed amplifier described previously.[60] A stepper motor (ThorLabs DRV50) was used to move the push rod until the bridge conductance was between 5 and 7 $G_0$, at which point a linear piezo actuator (ThorLabs PAZ60 or ThorLabs PAS009) was used to break and then re-form the bridge at a rate of 60 μm/s. The motor and the piezo were both controlled with custom LabView software that automatically collected thousands of breaking traces for each sample. The entire bending apparatus is built on a vibrationally isolated laser table to reduce mechanical noise, and placed inside a copper Faraday cage to reduce high-frequency electromagnetic noise.

**2.3. Molecular Solutions.** OPV2-2BT and all OPV3-2BT-X molecules were synthesized on-site, while C6-2SMe was purchased from Sigma-Millipore and used as received. OPV2-2BT and all OPV3-2BT-X molecules were dissolved in dichloromethane (HPLC grade, >99.8%), and C6-2SMe was dissolved in a mixture of hexanes (Reagent grade, >98.5%). All OPV3-2BT-X solutions were ~1 μM; both ~1 μM and ~10 μM solutions of OPV2-2BT were used (see SI Table S3 for details); the C6-2SMe solution was ~10 μM.

**2.4. Running Samples.** Each sample was cleaned with $O_3$/UV immediately before use and a Kalrez gasket (0.114" ID, 0.250" OD) was placed around the gold bridge (Fig. S1d). Initially, 10 μL of pure dichloromethane or hexanes was deposited inside this gasket using a clean glass syringe for dichloromethane or a micropipette for hexanes, after which a few thousand breaking traces were collected. Only samples displaying clean breaking and clear tunneling behavior were considered for subsequent experiments. After pausing the LabView program and fully breaking the gold bridge, 10-20 µL of the molecular solution was deposited inside the Kalrez gasket using a clean glass syringe for dichloromethane solutions or micropipette for hexanes solutions, and data acquisition was resumed. For many samples, molecular solution or pure solvent was re-deposited multiple times and/or the push rod was fully relaxed prior to restarting the experiment (see SI section S.4 for details).

**2.5. Initial Data Processing.** The voltage applied to the piezo actuator was converted to piezo displacement using a previously performed interferometric calibration. For each sample, the conversion factor between piezo displacement and inter-electrode distance was determined by fitting the distribution of tunneling slopes from the traces collected before molecular deposition (see SI section S.2 for details), and this conversion factor was applied to all traces collected with that sample. Each breaking trace was aligned at zero inter-electrode distance using its last crossing of 0.7 $G_0$ following the method of Mischenko *et al.*[61] Breaking traces with no data points between 0.8 and 1.2 $G_0$ were excluded from subsequent analysis (typically < 1% of total breaking traces).

3. RESULTS AND DISCUSSION

**3.1. Description of Segment Clustering.** *3.1.1. Motivation.* A key consideration when deciding how to cluster multidimensional data is what type of object to cluster. In the case of break junction distance-conductance traces, two natural choices are to treat each trace as a single object ("trace clustering", which most approaches[15,33–35,37–42,44] have used so far) or to treat different visited points in distance-conductance space as individual objects ("point clustering", which we used in a previously reported clustering approach[36]). Neither choice is inherently superior to the other. Instead, each has potential advantages that are best understood by considering the question of how much "history" distance-conductance traces have—*i.e.*, how much a trace's behavior at one distance is correlated with its behavior at a previous or future distance. If traces randomly transition between different stable distance/conductance configurations (*i.e.* traces have "no history"), then point clustering can better identify these stable configurations whereas trace clustering may get confused by the random trajectories. On the other hand, if trace trajectories are highly non-random (*i.e.* traces have "significant history"), then trace clustering can identify groupings of similar trajectories that point clustering will likely miss.

In our experience, however, real experimental traces fall somewhere between these extremes: they display "partial" or "local" history. To illustrate this, we calculate the correlation coefficient between the conductances of all traces at one specific distance with their conductances at a second distance. This is repeated for each pair of distances, and the results are summarized in a "distance correlation histogram", shown in Figure 2 for one of our OPV3-2BT-H datasets. This plot shows that while conductances are strongly correlated at close distances, there is essentially no correlation over longer distances. Similar behavior was found in all of the single molecule datasets we examined, suggesting that trace history is only relevant over short pieces of an entire trace. This is consistent with investigations of the dynamics of single-molecule junctions held at a fixed distance,[62–67] which have found that junction conductance is relatively stable over short periods of time, but jumps unpredictably between different levels over longer time windows. Therefore, both trace clustering and point clustering fail to fully and appropriately capture the empirical balance between predictable and random junction behaviors, limiting the insight they can provide. This motivates the development of a novel clustering approach in which *pieces of traces* are the type of object clustered.

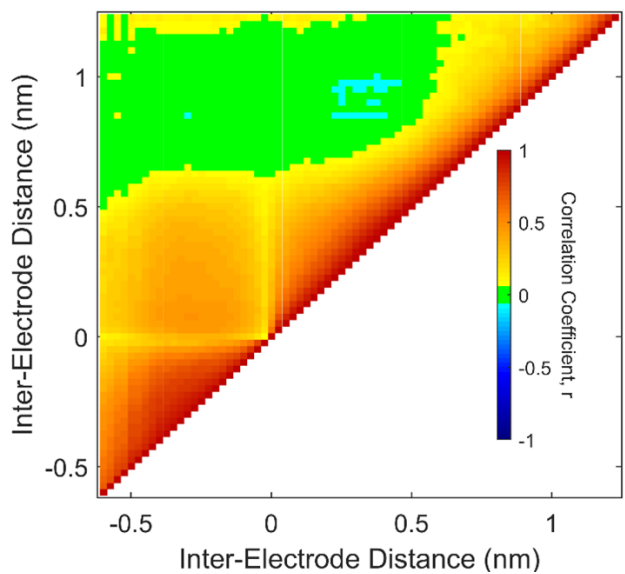

**Figure 2.** Distance correlation histogram for the OPV3-2BT-H dataset from Fig. 1c, showing the Pearson's correlation coefficient between the conductances of all traces at each pair of distances. While trace conductances are highly correlated over short distances, this correlation quickly fades with distance, demonstrating that trace "history" is important only locally, not globally.

While certain theoretical models predict significantly curved trace features, experimental traces collected from an extremely wide variety of molecular systems[13,51–59] appear (on a logarithmic conductance scale) to be composed mainly of sudden changes between fairly linear sections. Segment Clustering is therefore based on using piecewise-linear approximation to determine where to separate each trace into different sections. This design choice helps ignore noisy high frequency components and instead focuses attention on the principal features of each trace. Additionally, linear segments are a computationally efficient way to represent a trace, since a handful of linear segments can well-approximate thousands of individual data points (*e.g.* Fig. 3a). Implementing Segment Clustering *via* this approach consists of four major steps, summarized in Figure 3: segmentation, parameterization, calculating the overall clustering structure, and extracting specific clusters. Where appropriate, we employ established algorithms for these individual steps in order to increase

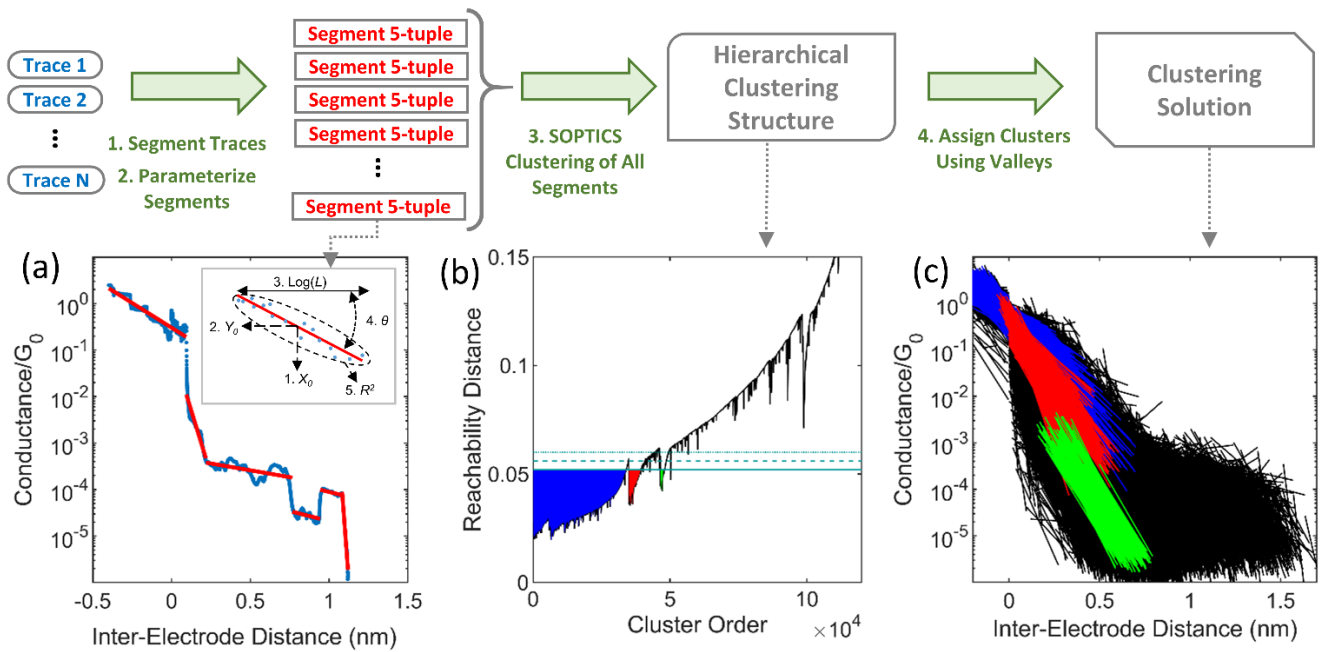

**Figure 3.** A summary of the segment clustering process. (a) Each breaking trace in a dataset is first approximated with a series of linear segments using BUS with the Greedy Iterative L-Method, and then each segment is parameterized to produce a 5-tuple (see inset). (b) Next, the set of 5-tuples for all segments from all traces in the dataset are clustered using the SOPTICS algorithm, producing a hierarchical clustering structure that can be visualized using a reachability plot in which valleys correspond to clusters. Finally, a specific clustering solution can be extracted by making a cut through the reachability plot and assigning the points in each valley dipping below that level to a separate cluster, while assigning any points with reachability distances greater than the cut to a catch-all "noise cluster". Extracting at the solid blue line in (b) produces the clustering solution in (c), with each valley dipping below the line filled in with color to match its corresponding cluster of segments, and the noise cluster segments shown in black.

confidence in the robustness of the overall approach, which combines these algorithms in a new way.

*3.1.2. Segmentation.* The goal of segmentation is to break each trace into consecutive sections such that each section can be well-represented by a linear segment and corresponds to a meaningful piece of the trace structure. Because this goal is common in data-mining applications, several algorithms have been developed to try to optimally represent time-series data with a set of piece-wise linear segments.[68] After first applying consistent starting and ending criteria to each trace (see SI section S.3.1 for details), we employ the "Bottom-Up Segmentation" (BUS) algorithm because it is conceptually simple and has been found to produce excellent and robust results for data from a variety of contexts.[68,69] Briefly, BUS starts by perfectly representing a time series of $n$ points with $n/2$ two-point segments. Next, BUS iteratively merges the pair of neighboring segments that will least increase the error of the overall segment approximation, repeating until some stopping criteria is met. At each step, every segment is constructed as the linear regression line for the data points it is currently representing, and the error for each segment is taken as the sum of the squared residuals from that regression line.[68]

For our stopping criteria, we use the "Greedy Iterative L-Method", which was found to work well on a wide variety of test datasets.[69] Briefly, this method first performs the merging process to completion, so that a plot of the number of segments remaining vs. the error gained at each merge step may be constructed. An iterative fitting process is then used to locate the optimal number of segments by identifying the point at which more segments produce diminishing returns in terms of error reduction. Applying this combination of BUS and the Greedy Iterative L-Method to distance-log(conductance) traces produces convincing segmentation solutions (*e.g.* Fig. 3a). In addition to the examples presented by the developers of the Greedy Iterative L-Method,[69] testing on our own single molecule data demonstrates that this method is quite robust (see SI section S.5.5).

*3.1.3. Parameterization.* Because clustering algorithms need to compute distances between the objects to be clustered, it is necessary to first extract "features" that can be used to represent each object as a point in a metric space. In order to avoid well-known challenges to clustering in high-dimensional spaces (the "curse of dimensionality")—such as increasingly sparse data and a non-intuitive breakdown in the concept of nearest neighbors[70]—it is preferable to choose a minimal set of features while still capturing most of the important information about each trace piece. Our segmentation approach already produces linear segments which capture most trace variation—*e.g.*, 82% for the dataset in Figure 1c—and so parameterizing these linear segments produces features that are both efficient and easy to interpret. We therefore convert each segment into a 5-tuple consisting of four parameters which uniquely describe each linear fit, and a fifth parameter to describe the fit quality.

The specific parameters chosen to represent each segment are illustrated in Figure 3a. The first two parameters—the center of a segment on the inter-electrode distance axis, $X_0$, and on the log(conductance) axis, $Y_0$—succinctly represent where each segment is located. Another key segment attribute is its length, $L$. However, in absolute terms, long segments will tend to differ by more than short ones, making it difficult to form clusters of long segments. We therefore use the logarithm of the length of a segment on the inter-electrode distance axis, Log($L$), as our third parameter, so that the difference between two segments on this dimension depends on their ratio. To represent how tilted a segment is, the angle that it makes with the horizontal, $\theta$, is used as the fourth parameter. This angle is less sensitive to outliers than a segment's raw slope due to the nature of the *arctan* function. Finally, to represent the linearity of each trace piece, we include the coefficient of determination, $R^2$, of each segment vis-à-vis the portion of raw data it represents as the fifth parameter. This helps capture additional information about mild segment curvature and/or the magnitude of high-frequency noise, and is important for differentiating the few segments which are not well-approximated as linear. These five parameters are each measured in different units, so before clustering each must be standardized so that differences computed along different dimensions are comparable. In order to minimize the influence of outliers, we use the range of the middle 80% for each parameter to carry out this standardization (see SI section S.3.2 for details).

*3.1.4. Calculating the Overall Clustering Structure.* Many clustering algorithms can be applied to a set of 5-tuples, and each has its own advantages and disadvantages.[71] For this work, we employ the Ordering Points to Infer Cluster Structure (OPTICS) algorithm based on the following advantages relevant to our specific context: 1) it can detect clusters of arbitrary shape and is not biased towards spherical clusters like other common algorithms;[71,72] we acknowledge that this necessarily brings along a danger that dissimilar groups of data may end up in the same cluster if there is a continuous spread of data between them; 2) it has a limited number of parameters; 3) it does not require the number of clusters to be specified as an input parameter, unlike many popular algorithms such as K-means, BIRCH, *etc.*;[72] and 4) instead of a single partitioning, OPTICS produces a clustering hierarchy in which sub-clusters are contained within clusters, providing relevant insight into the data structure (see below). To overcome its poor computational scalability on large datasets, we employ a variation called Speedy-OPTICS (SOPTICS) in which random projections are used to dramatically reduce the clustering time while producing results nearly identical to the original algorithm.[73]

OPTICS/SOPTICS clustering works by starting at a random data point, then iteratively proceeding to the next unvisited point that is closest to any point visited so far.[36,74] This journey is represented by a "reachability plot" (Fig. 3b) in which the distance to the next point (the "reachability distance") is plotted against the order in which the points were visited (the "cluster order"). *Valleys* in the reachability plot intuitively correspond to *clusters* of data points, because the points in a valley are relatively close to each other but relatively far from points outside of the valley.[74] A reachability plot thus visually represents the overall hierarchical structure of a dataset, as valleys may contain sub-valleys which themselves can contain sub-sub-valleys, and so on. We refer to the reachability plot and its associated information as the "clustering output" for a given dataset.

In our implementation, SOPTICS relies on four parameters: $c_L$, $c_P$, *minSize*, and *minPts*. The first three parameters are related to how SOPTICS approximates the original OPTICS algorithm and, when in a reasonable range, they each have an extremely minimal effect on the clustering results. We thus assign fixed values to each of these parameters (see SI section S.3.3 for details). The fourth parameter, *minPts*, is the one holdover from OPTICS (SOPTICS does not require the generating distance parameter ε); it is related to how the data density in 5-dimensional standardized parameter space is estimated at each point, and affects how "jagged" the reachability plot is.[74] While *minPts* is the most important parameter for OPTICS/SOPTICS, its abstract definition makes it difficult to assign rationally without a deep understanding of the data under consideration. In acknowledgement of this uncertainty, we re-cluster each dataset using 12 different values of *minPts* (35, 45, 55, 65, 75, 85, 95, 105, 115, 125, 135, and 145). We then use the variation between these 12 clustering outputs as a measure of the uncertainty in the exact boundaries of an extracted cluster. In practice, this variation is quite limited, implying that Segment Clustering is not overly sensitive to the value of *minPts*. Finally, because OPTICS/SOPTICS is a density-based clustering algorithm and longer segments represent more raw data points than shorter segments, we find that clustering results are improved if, in the density calculations, we weight each segment according to its length (see SI section S.3.4 for details).

*3.1.5. Extracting Specific Clusters.* In order to extract specific clusters from a given clustering output, a cut is made across the reachability plot (*e.g.* Fig 3b) and the points in each valley dipping below the cut are assigned to a separate cluster, while all points with reachability distances larger than the cut are assigned to a catch-all "noise cluster" (*e.g.* Fig 3c). We refer to the specific set of clusters generated by a given cut as a "clustering solution". Thus, while the hierarchical nature of OPTICS/SOPTICS is a distinct advantage, it also presents an interpretation challenge, because a single clustering *output* can have many different clustering *solutions* based on different extraction levels.

Meaningful extraction levels can be chosen using the concept of ξ-steepness[74] or by employing an internal cluster validation index,[75,76] but these strategies introduce ambiguity in the form of what value of ξ to use or which index to employ, and many validation indices are expensive to compute. We therefore introduce a new strategy motivated by the observation that the clustering solutions at most extraction levels are extremely similar to one another. For example, Figure 3c shows the clustering solution obtained by extracting at the solid line in Figure 3b. If this extraction level is increased to the dashed line, the only change is that each valley grows slightly, with a few segments moving into those clusters from the noise cluster. The clustering solution will only *qualitatively* change if the extraction level is raised, for example, to the dotted line, where the red and blue valleys/clusters will merge into one. In the context of Segment Clustering, we are interested in categorizing as many data points as possible, so we extract each individual valley at the highest extraction level before it merges with a neighboring valley to produce what we call "maximum valley clusters". If a minimum valley size is then set, an entire clustering output can be efficiently summarized with just a handful of maximum valley clusters (Fig. 4). This allows us to still examine the hierarchical structure of a clustering output without having to consider an unmanageable number of different solutions. This novel extraction strategy works especially well in the present context because valleys tend to be quite sharp (e.g. Fig. 4a), and its robustness is validated by the fact that it successfully identifies equivalent clusters in the multiple clustering outputs for each dataset (see SI section S.6 for details). However, we note that this extraction approach is not fundamental to Segment Clustering, and so other methods can be substituted if full-valley clusters were to exhibit shortcomings on new types of datasets. The minimum valley size should be set according to the specific context and what types of clusters a user is interested in; we have found that a minimum size of 1% of the total number of data points (after length-weighting) often works well.

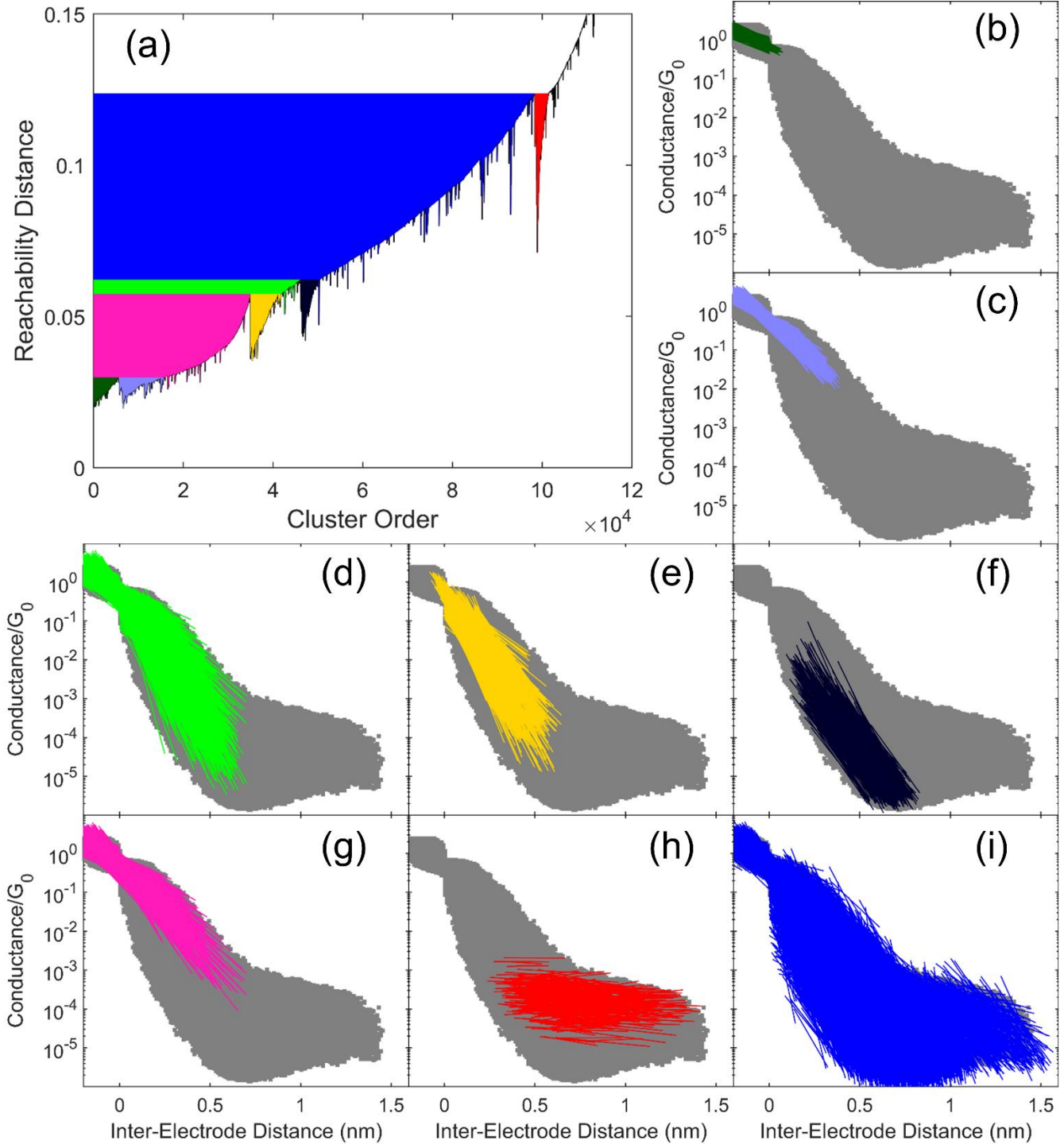

**Figure 4.** (a) The same reachability plot as in Fig. 3b, but color-coded to indicate the maximum size of each valley containing at least 1% of all clustered points. Valleys are filled in hierarchically: the pink valley *e.g.* contains the dark green and lavender valleys, the green valley contains the pink and yellow valleys, *etc.* (b-i) The "full-valley clusters" corresponding to each color-coded valley from (a), with segments assigned to the cluster plotted in color on top of the overall dataset distribution in gray.

**3.2. Using Segment Clustering to Distinguish the Conductances of Similar Molecules.** In structure-property investigations of single molecule conductance, it is common to determine a single "most probable" conductance for each molecule by fitting the molecular peak in the 1D histogram.[17–20,22,24,26] The peak value is then identified as the molecular conductance, and often compared across different molecules or with first principles calculations. However, because the molecular signal is necessarily convolved with a "background" signal due to traces in which no molecule was bound or in which the molecule detaches and re-attaches multiples times (*e.g.* Fig. 1a), molecular peaks in 1D histograms tend to have complex, asymmetric shapes (*e.g.* Fig. 1d). Fitting these peaks thus requires arbitrary and ill-motivated restrictions and/or background subtraction. Moreover, it has been shown that the molecular peak can vary significantly between repeated measurements under identical conditions,[33] likely due in part to uncontrolled variation of this "background" signal. Using data collected from a series of OPV3-2BT-X molecules (Chart 1), we show how segment clustering can help address these twin challenges by separating the primary molecular feature from the background signal, enabling subtle conductance differences to be identified with confidence.

*3.2.1. Extraction of "Main Plateau Cluster" From Background.* In order to perform this background separation, we examined each full-valley cluster for the OPV3-2BT-H dataset shown in Figure 4b-i. Of these, the red cluster (Fig. 4h)

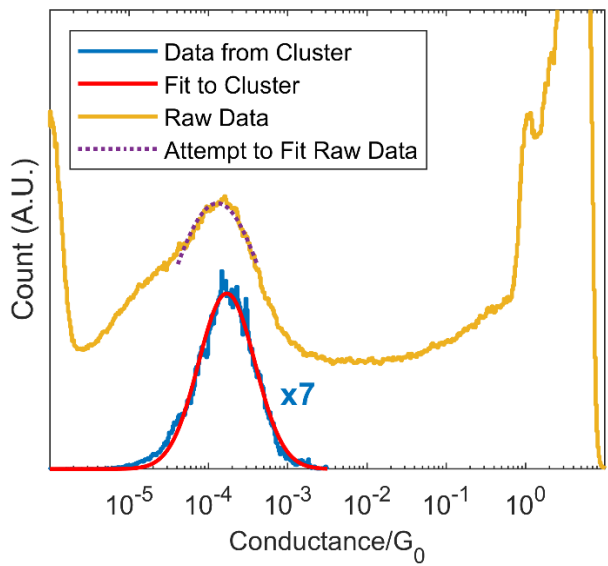

**Figure 5.** Raw 1D histogram for the OPV3-2BT-H dataset from Fig. 1c (yellow), along with a restricted Gaussian fit to the molecular peak (dotted purple, see SI section S.8 for details). Overlaid in blue is a 1D histogram of the data from just the main plateau cluster (Fig. 4h) and an unrestricted Gaussian fit (red), both scaled up by a factor of seven for clarity. Whereas the complex shape of the raw data peak necessitates arbitrary fitting restrictions to obtain reasonable results, the simple shape of the main plateau cluster peak can be fit without restrictions, leading to a more confident and robust peak value.

is the unambiguous choice for the primary molecular signature because 1) it most closely corresponds to the dense molecular region in Fig. 1c that is not present in Fig. 1b, and 2) it is composed of relatively long and flat molecular plateaus that approximately match the expected length of the molecule after adding 0.5 nm to account for the "snapback" distance[28,77] (see SI section S.9 for details). We therefore refer to the cluster in Fig. 4h as the "main plateau cluster". In contrast to the raw data, the conductance peak for the main plateau cluster has a simple shape that can be confidently fit with no restrictions by a single Gaussian (Fig. 5). This is a direct consequence of Segment Clustering's novel focus on pieces of traces as the clustering unit, since trace clustering approaches will necessarily produce clusters with complex conductance histogram shapes. However, the main plateau cluster in Fig. 4h does not represent *all* of the molecular signature in the dataset. In fact, the points in these segments only account for a small fraction of the molecular peak seen in the raw 1D histogram (Fig. 5). This may be caused by a majority of molecular traces at room temperature jumping back and forth between tunneling decay and molecular plateaus (*e.g.* Fig. 1a), whereas the segments in the main plateau cluster only originate from the "cleanest" molecular plateaus (*i.e.* those that are long, unbroken, and relatively constant). We hypothesize that these "cleanest" plateaus will yield the most reliable measure of molecular conductance and the underlying

To test this hypothesis, we collected nine total OPV3-2BT-H datasets across three different samples run under identical conditions (see SI section S.4 for details). Within all but one of these datasets (see SI section S.7 for details), a main plateau cluster analogous to the one shown in Figure 4h could be unambiguously identified (Fig. 6a-h), providing strong evidence that this type of cluster is a meaningful and reproducible structural element of these datasets. Each of these main plateau clusters can again be effectively fit with an unrestricted single Gaussian (see SI section S.8 for details). Comparing the spread of these 8 peaks with the restricted peaks fit to the raw 1D histograms (Fig. 6i) reveals a significantly tightened distribution (Table 1), consistent with our hypothesis that segment clustering is aiding the extraction of an inherent molecular feature from a widely varying background.

**Table 1.** Comparison of different measures of spread for the raw data peaks vs. the main plateau cluster peaks for 8 different OPV3-2BT-H datasets (see Fig. 6i), demonstrating that Segment Clustering increases the reproducibility of peak conductance measurements. All units are decades.

|  | Raw Data Peaks | Main Plateau Cluster Peaks |
|---|---|---|
| Range | 0.159 | 0.099 |
| Standard Deviation | 0.063 | 0.032 |
| Inter-Quartile Range | 0.121 | 0.037 |

*3.2.2. Quantitative Comparison of Conductances of Similar Molecules.* Figures 5 and 6 demonstrate the power of segment clustering: the need for complex and arbitrary fitting criteria is eliminated *and* dataset-to-dataset reproducibility is improved, allowing us to identify peak molecular conductances with increased precision and confidence. To illustrate the advantages of this increased precision, we used our MCBJ set-up to collect multiple sets of breaking traces for a total of seven OPV3-2BT-X molecules (Chart 1; see SI section S.4 for details on datasets). For all but two datasets (see SI section S.7 for details), we identified a clear and unambiguous choice for the full-valley cluster corresponding to the main plateau feature. Our peak conductance results for all of these OPV3-2BT-X main plateau clusters are summarized in Figure 7, in which the error bars represent the uncertainty introduced by varying the *minPts* parameter (see SI section S.6 for details).

Figure 7 shows that, as with OPV3-2BT-H, the peak conductances for each molecule in the series are highly reproducible, further supporting the claim that segment clustering is extracting an inherently molecular feature. Moreover, because of this high reproducibility, we are able to confidently *differentiate* the conductances of these molecules despite their high structural similarity. This makes it possible to search for structure-property relationships to physically quantum transport, which is otherwise obscured by the large and stochastically visited space of possible junction configurations.

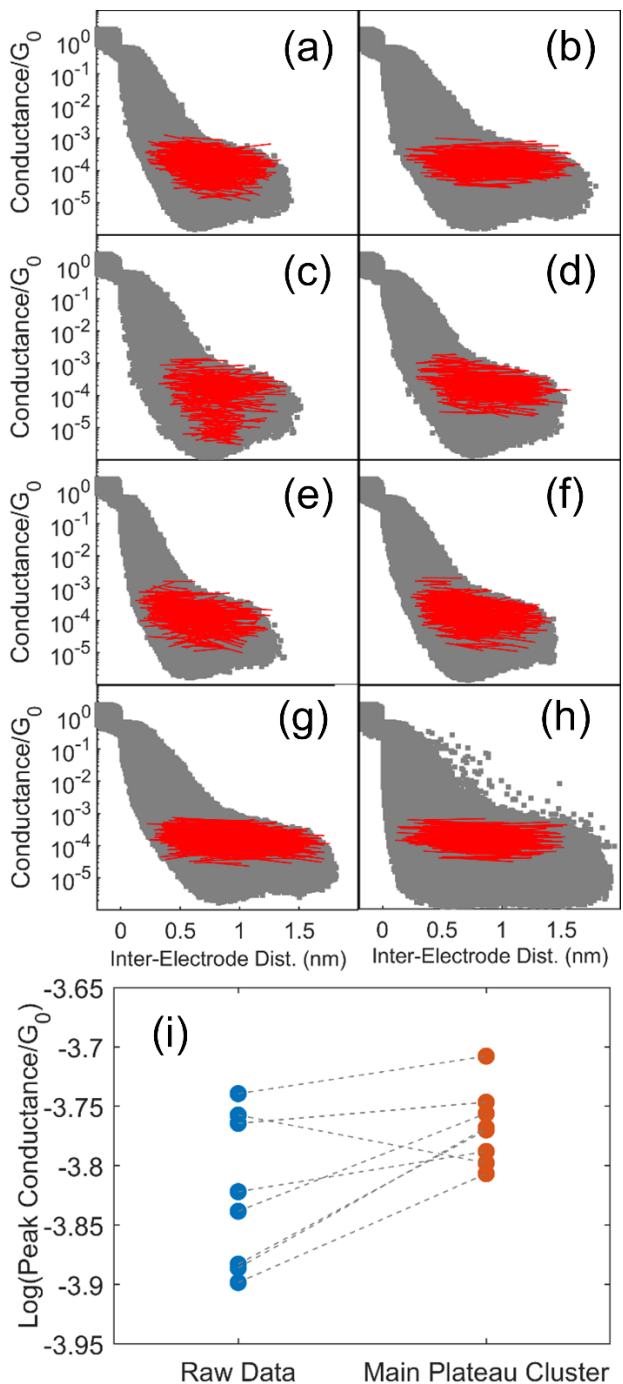

**Figure 6.** (a-h) Main plateau clusters selected for 8 different OPV3-2BT-H datasets, demonstrating that this feature is a consistent structural element of these datasets. (i) Comparison of peak conductance values from unrestricted Gaussian fits to the main plateau clusters from (a-h) with the peak conductance values from restricted Gaussian fits to the raw 1D histograms (see SI section S.8 for details), demonstrating that Segment Clustering increases the precision of peak conductance measurements.

explain such conductance differences. Extensive testing confirms that the peak conductances in Figure 7 are not meaningfully affected qualitatively *or quantitatively* by modest changes to the clustering parameters (see SI section S.5 for details). Not only does this increase confidence in these specific results, but it also provides strong evidence that Segment Clustering is a highly-robust and generalizable tool for unsupervised analysis of potentially subtle variations in molecular conductances.

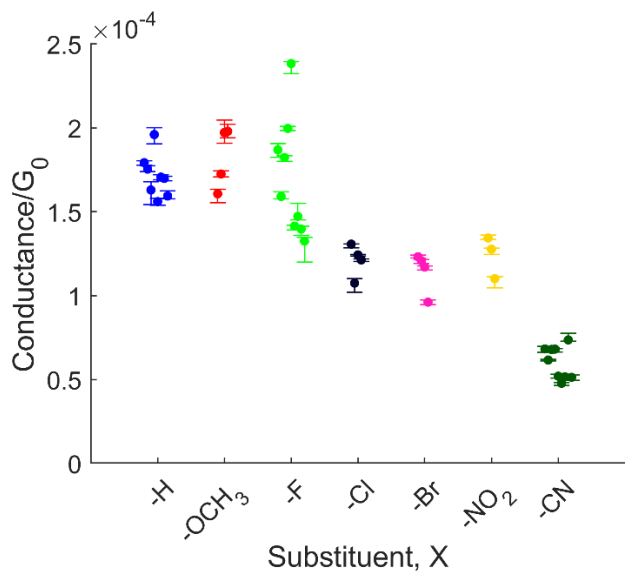

**Figure 7.** Comparison of peak conductance values from main plateau clusters for each OPV3-2BT-X dataset considered in this work. Error bars represent the uncertainty due to clustering with different values of the *minPts* parameter (see SI section S.6 for details). Due to the high reproducibility enabled by Segment Clustering, subtle conductance differences between molecules can be identified with confidence.

**3.3. Using Segment Clustering to Separate Overlapping Molecular Features.** In addition to the extraction of a single "primary" molecular feature in different data sets, Segment Clustering can also be used to distinguish multiple features in a single dataset. When 2D histograms of breaking traces display multiple "clouds" of increased density, it is often taken as an impetus to investigate the possibility of different binding modes, molecular configurations, *etc*.[9,28–32] While such clouds can offer tantalizing hints of multiple transport motifs, a major challenge is that it is often quite ambiguous whether density clouds are truly separate or not. This introduces a significant opportunity for bias, and may also limit the scope of hypotheses considered for further investigation. Because segment clustering is unsupervised and largely model-free, it is a useful tool for *objective* separation of molecular features.

To demonstrate this, we constructed a synthetic dataset consisting of equal numbers of experimental traces from samples run with two structurally rather different molecules. The first half of traces are taken from a dataset collected with the molecule C6-2SMe (Chart 1), which displays a short molecular feature at ~$10^{-4}$ $G_0$ (Fig. 8a,c). Segment clustering of this dataset unambiguously identified a full-valley cluster corresponding to this molecular feature (Fig. 8b,c; see SI section S.10 for details). The remaining traces for our synthetic mixture are taken from an OPV2-2BT (Chart 1) dataset. The histograms of the breaking traces for this molecule reveal a strong high-conductance feature at ~$10^{-3}$ $G_0$ as well as a subtler low-conductance feature at ~$10^{-4}$ $G_0$ (Fig. 8d,f), likely due to molecular stacking or direct π-Au binding.[21,59] While segment clustering identifies a main plateau cluster corresponding to the high-conductance feature (Fig. 8e), none of the full-valley clusters matches well with the low-conductance feature (see SI section S.10 for details). This shows that segment clustering will not always extract every meaningful feature from a dataset.

However, because the low-conductance feature of OPV2-2BT partially overlaps the primary C6-2SMe feature, our synthetic mixture provides an excellent challenge case for Segment Clustering. This can be seen in the 2D histogram for our mixture (Fig. 8g), which is qualitatively quite similar to the pure OPV2-2BT histogram (Fig. 8d) and displays exactly the type of ambiguous dual density cloud often reported in literature,[9,28–31] and sometimes imbued with speculative microscopic meaning. Moreover, Figure 8i shows that the intensity and location of the lower peak in the 1D histogram of our synthetic mixture falls *within* the variability observed between different pure OPV2-2BT datasets, further illustrating the challenge posed by separating these two molecular distributions.

As shown in Figure 8h, segment clustering of our mixture dataset identifies two full-valley clusters that appear to correspond to the main OPV2-2BT and C6-2SMe features (though because both molecular features are "diluted" by mixing, the minimum valley size was lowered below 1% to locate these valleys; see SI section S.11 for details). Because this mixture was constructed synthetically, we can quantitatively test this hypothesis. We find that the separation of molecular features is indeed quite accurate, even though the two clusters partially overlap: 97% of the data in the OPV2-2BT cluster belong to traces taken from the OPV2-2BT dataset, and 84% of the data in the C6-2SMe cluster come from C6-2SMe traces. It is not surprising that the C6-2SMe cluster has a higher misidentification rate, because this cluster's shorter segments are much more likely to be found in an arbitrary dataset simply by chance. This is evidenced by the fact that a *cluster* of C6-2SMe-like segments *did not exist* in the pure OPV2-2BT dataset, indicating that the misassigned segments added to the C6-2SMe cluster from the mixture dataset did not form a region of high density by themselves. To further test the robustness of this feature separation, we constructed seven additional 1:1 OPV2-2BT:C6-2SMe synthetic mixture datasets using different combinations of traces from different pure-molecule datasets (see SI section S.4 for details). As shown in Figure S17 and Table S6 in the SI, segment clustering successfully extracted both molecular features for all but one of these mixtures (see SI section S.12 for details), and each of these separations displayed high quantitative accuracy.

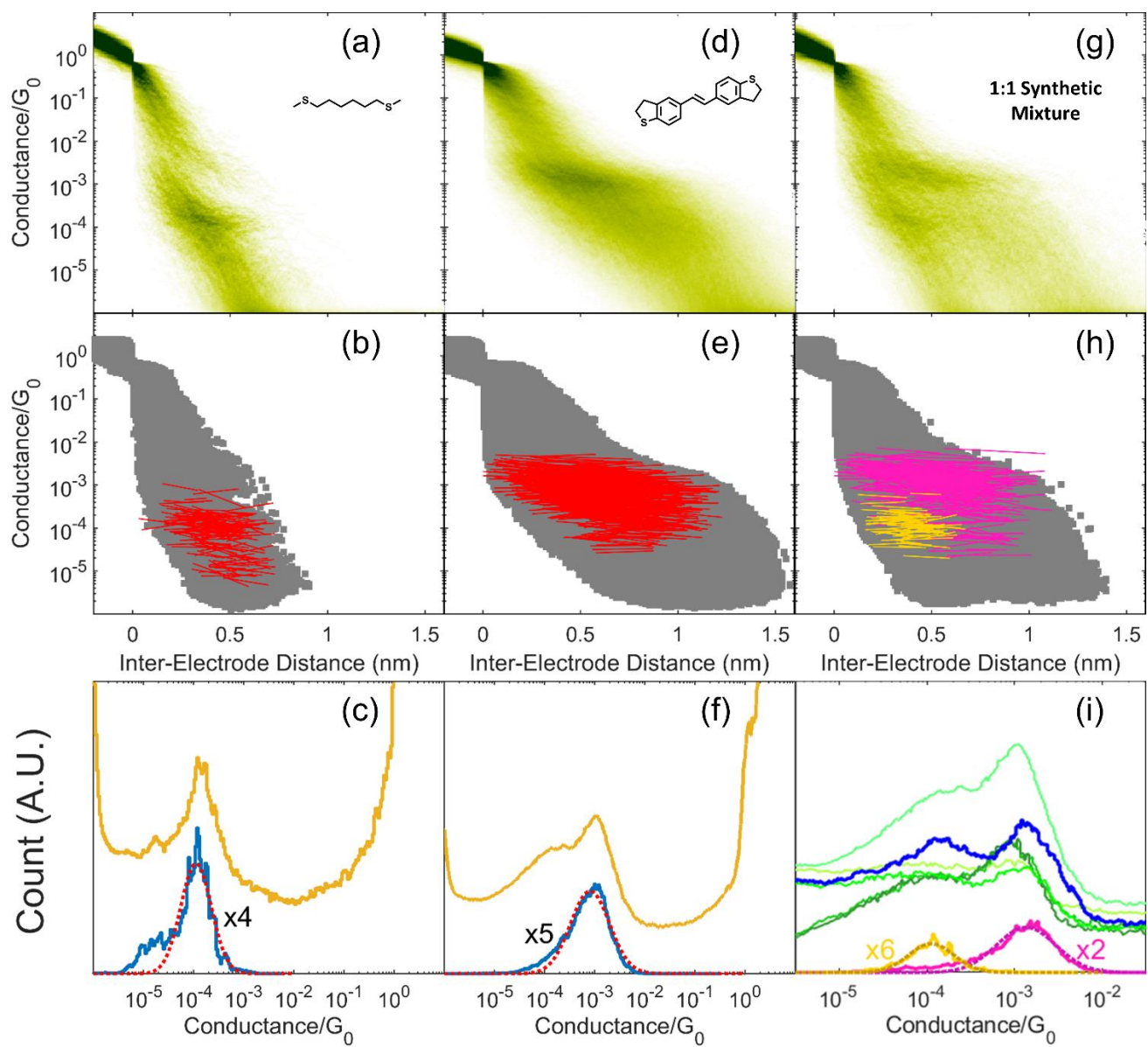

**Figure 8.** (a) 2D histogram for 1315 consecutive breaking traces collected in the presence of C6-2SMe. (b) The full-valley cluster identified as the main plateau cluster for the data from (a). (c) 1D histogram for the raw data from (a) (yellow), overlaid with the 1D histogram for the data from the main plateau cluster in (b) (blue) with an unrestricted Gaussian fit (dotted red). (d-f) Analogous plots to (a-c) for a dataset containing 5807 consecutive breaking traces collected in the presence of OPV2-2BT. (g) 2D histogram for a synthetic dataset constructed by combining equal numbers of traces from the datasets in (a) and (d). (h) The two full-valley clusters identified as molecular plateau features for the data from (g). (i) 1D histogram for the data from (g) (dark blue), overlaid with the 1D histograms for the two clusters from (h) (pink and yellow) and their respective single Gaussian fits (dotted lines). For comparison, 1D histograms for 5 different raw OPV2-2BT datasets are included (various shades of green), demonstrating that the intensity and location of the peaks in the synthetic mixture lie well within the range of the different pure OPV2-2BT datasets.

By reliably separating features in an experimental dataset, Segment Clustering contributes to an important goal of single molecule transport research, towards which some progress has already been made. For example, several existing clustering algorithms have a demonstrated ability to extract multiple sub-features from experimental datasets of one molecular species.[33,34,36,38,42] However, while these studies offer intriguing hints about different binding modes and molecular conformations, such sub-features are unfortunately difficult to corroborate without extremely trustworthy atomistic simulations. More-testable examples of feature separation have been demonstrated by Hamill *et al.*, whose sorting algorithm successfully separated the features for two molecules in a mixture displaying an "obvious bimodal feature",[34] and recently by Huang *et al.*, whose deep-learning clustering algorithm separated two features from an overlapping molecular mixture.[37] However, because neither mixture was synthetic, these separations could not be quantitatively confirmed for the accuracy of cluster assignments. Finally, Vladyka and Albrecht very recently applied a neural network-based classification algorithm to a synthetic mixture of three different molecules, and while some pairwise separation was qualitatively observed, the combination of all three molecular features could not be separated.[41] The OPV2-2BT/C6-2SMe case study described here is thus a significant advance in that it constitutes a quantitatively validated example of experimental feature separation, and it does so in the challenging case of overlapping features. This provides a powerful demonstration of the usefulness of Segment Clustering as a hypothesis generation tool.

## 4. CONCLUSIONS

In this work we presented Segment Clustering, a novel approach to aid hypothesis generation for datasets of single-molecule breaking traces. Segment Clustering is *categorically* different than all previous clustering approaches since it treats, for the first time, pieces of breaking traces as the fundamental clustering unit, allowing behaviors occurring in just part of a trace to be more readily identified. This sub-trace focus gives Segment Clustering the potential to yield new and powerful insights into single-molecule datasets because grouping the data by segments is a better match for the empirical "local history" and piece-wise linear structure of break junction data than grouping by entire traces. This suggests that the segmentation approach described here may be a valuable avenue for future investigations even outside the context of clustering, for example by comparing the distribution of segment lengths between different datasets or exploring the likelihood of certain types of segments to appear in the same traces as others. To encourage such new directions, and to enable the use of the Segment Clustering in other contexts, we have made our code freely available in a user-friendly open-source package (github.com/LabMonti/SMAUG-Toolbox).

To demonstrate the power and versatility of the full Segment Clustering approach, we have applied it to two common challenges faced in the analysis of breaking traces. First, to address the related issues of complex peak shapes and varying background signals in conductance histograms, we used Segment Clustering to extract the "primary" molecular feature in a series of similar molecules. We showed that this increases measurement reproducibility *and* the robustness of peak-fitting, allowing subtle conductance changes to be distinguished with confidence. Second, to address the problem of separating ambiguous or overlapping molecular features, we used Segment Clustering to search for clusters corresponding to particular features in 1D and 2D histograms. By constructing a synthetic mixture of traces from two different molecules with overlapping conductance distributions, we demonstrated that Segment Clustering performs this feature separation with high quantitative accuracy even in challenging circumstances. We expect that these two advances in particular, as well as the new perspective offered by Segment Clustering in general, will aid in the establishment of structure-property relationships in single molecule quantum transport and thus help unlock new paths toward harnessing molecular electronics by design.

## ASSOCIATED CONTENT

**Supporting information:** MCBJ set-up; Inter-electrode distance calibration; Additional design criteria for Segment Clustering; Dataset collection and construction; Robustness of OPV3-2BT-X results to clustering parameters; Selecting clusters from multiple clustering outputs for the same dataset; Selection of main plateau clusters for OPV3-2BT-X datasets; Peak fitting; Investigating main plateau cluster lengths; Selection of main plateau clusters for OPV2-2BT and C6-2SMe; Cluster selection for OPV2-2BT/C6-2SMe 1:1 synthetic mixture #1; and Clustering of additional synthetic mixtures. This material is available free of charge *via* the Internet at http://pubs.acs.org.

The MatLab code used for this work is available free of charge at github.com/LabMonti/SMAUG-Toolbox.

## AUTHOR INFORMATION

### Corresponding Author

*Oliver L.A. Monti: Department of Chemistry and Biochemistry, University of Arizona, Tucson, Arizona 85721, United States; Department of Physics, University of Arizona, Tucson, Arizona 85721, United States; Email: monti@u.arizona.edu; Phone: ++ 520 626 1177; orcid.org/0000-0002-0974-7253.

### Author Contributions

O.M., J.I., and N.B. conceived the research ideas. K.P. synthesized the OPV2-2BT and OPV3-2BT-X molecules, directed by D.M. MCBJ samples were fabricated and run by J.I. and N.B. Segment clustering was developed and implemented by N.B. with advice and input from O.M. and J.I. N.B. wrote the manuscript with input and advice from all authors.

### Notes

The authors declare no competing financial interests.

## ACKNOWLEDGEMENTS

The authors would like to acknowledge support from the National Science Foundation award no. DMR-1708443, as well as from the Graduate and Professional Student Council at The University of Arizona. Plasma etching was performed in part using a Plasmatherm reactive ion etcher acquired through an NSF MRI grant, award no. ECCS-1725571. Clustering was performed using High Performance Computing (HPC) resources supported by the University of Arizona TRIF, UITS, and RDI and maintained by the UA Research Technologies department. All SEM images and data were collected in the W.M. Keck Center for Nano-Scale Imaging in the Department of Chemistry and Biochemistry at the University of Arizona with funding from the W.M. Keck Foundation Grant. The authors would also like to thank R. Himmelhuber for help and advice on sample fabrication, A. Garlant for help with laser interferometry, and C. Raithel and D. Dyer for feedback on the manuscript.

# Supplementary Information for: Unsupervised Segmentation-Based Machine Learning as an Advanced Analysis Tool for Single Molecule Break Junction Data

Nathan D. Bamberger[†], Jeffrey A. Ivie[†,‡], Keshaba N. Parida[†], Dominic V. McGrath[†], and Oliver L.A. Monti[†,§,]*

[†]Department of Chemistry and Biochemistry, University of Arizona, 1306 E. University Blvd., Tucson, Arizona 85721, USA

[§]Department of Physics, University of Arizona, 1118 E. Fourth Street, Tucson, Arizona 85721, USA

**Table of Contents**

*S.1 MCBJ Set-Up*

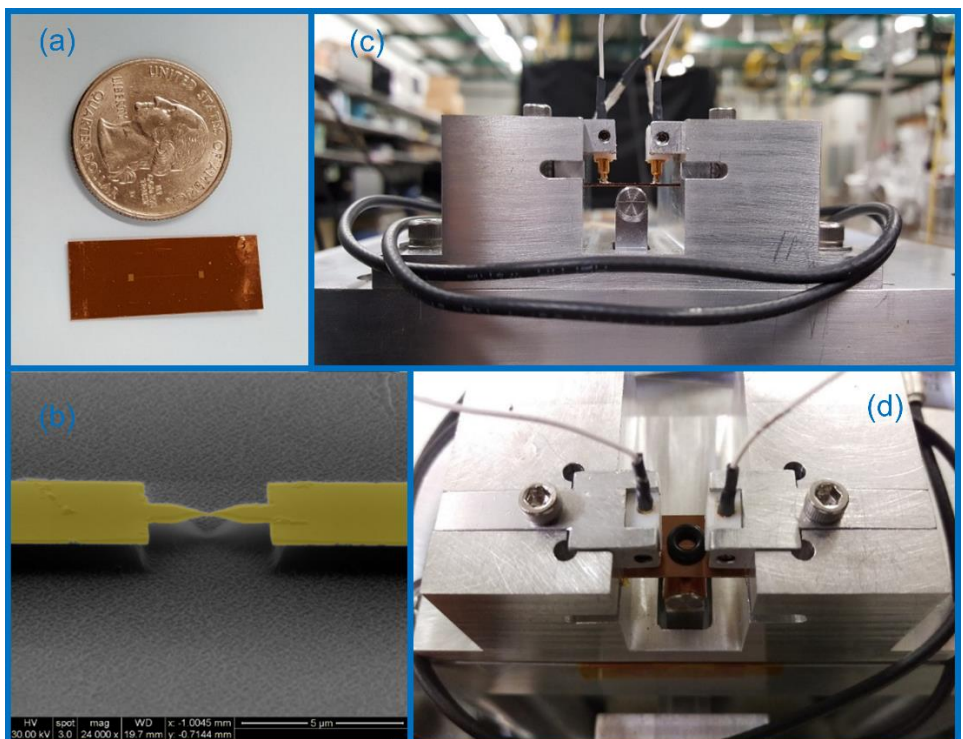

**Figure S1.** Images of MCBJ set-up. (a) Example of a lithographically defined MCBJ sample. (b) False color SEM image at 45° showing the suspended gold bridge in the center of a sample. (c) Side view of bending apparatus showing clamped-in sample with push rod underneath. (d) Top view of a clamped-in sample showing a Kalrez gasket placed around the center of the junction.

*S.2 Inter-Electrode Distance Calibration*

In an MCBJ set-up, the amount by which the two nano-electrodes pull apart (inter-electrode distance, $\Delta x$) for a given vertical movement of the push rod (piezo distance, $\Delta z$) is given by the "attenuation ratio" ($r = Δx/Δz$). While $r$ can be calculated using a simple model of elastic bending, this result tends to be wrong by a factor of 2 to 4 due to the inhomogeneous elastic properties of real lithographically defined junctions.[1] It is therefore preferable to experimentally determine the attenuation ratio via one of several possible calibration methods.[2] Because the attenuation ratio depends on the exact length of the suspended gold bridge, which varies from sample to sample, we independently calibrated $r$ for each sample considered in this work. To enable this calibration, each sample was run "bare" (no molecules deposited, only pure solvent which quickly evaporates) for a few thousand traces, and the $r$ value calculated from these traces was then applied to all subsequent traces collected with that sample.

For the calibration itself, we employ the method of the tunneling slope. For small bias voltages, the tunneling current between two nano-electrodes as a function of their separation, $x$, is well-approximated by $I(x) = I_0 \exp(-Bx)$, where $B$ is a constant depending on the effective work function of the electrodes.[3] A plot of $Log_{10}(G/G_0)$ vs. distance should therefore have a constant slope. By comparison to an STM-BJ set-up, Hong et al. found that this slope is 5.5 to 6 decades/nm for breaking traces collected in argon.[4] In an independent study, Grüter et al. found that the tunneling slope is ~1.7 times smaller for traces collected in air compared to in vacuum.[5] Based on high-quality data collected under vacuum,[2] this implies traces collected in air should have a tunneling slope of ~6 decades/nm, in agreement with the Hong et al. result, and thus we use this value for our calibration.

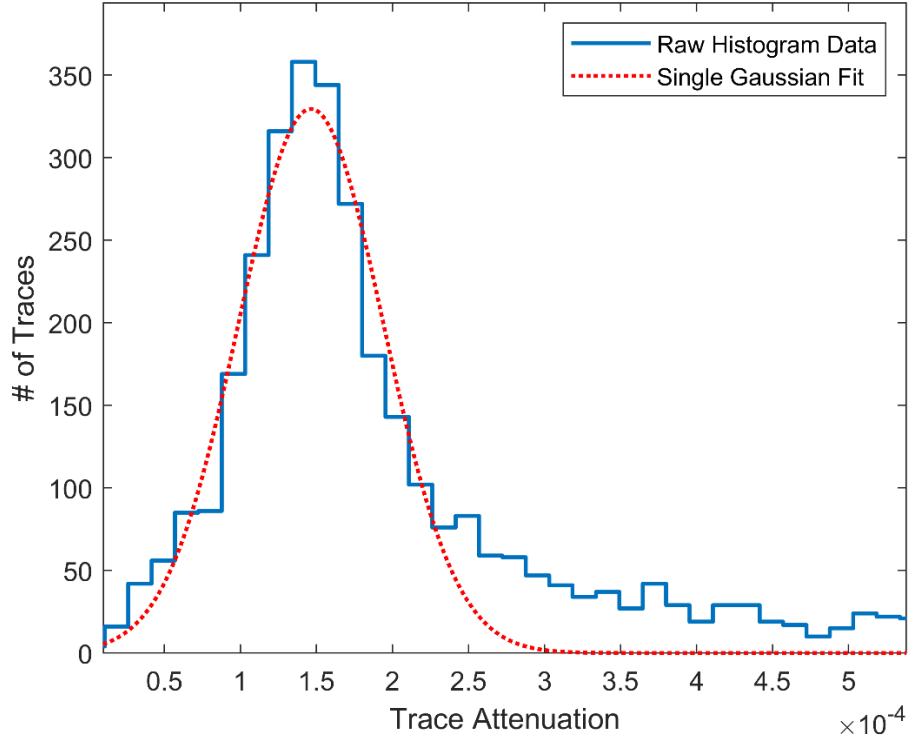

**Figure S2.** Histogram of 3481 individual trace attenuations (blue) and Gaussian fit (dotted red) used to determine the attenuation ratio for sample 113-2. Bin width was set to $1.54\times10^{-5}$ using the Freedman–Diaconis rule.

To perform the calibration, we first linearly fit the portion of each breaking trace below $2\times10^{-4}$ $G_0$, since this ensures that the tunneling slope is reliable,[2] and above $10^{-5}$ $G_0$, to be comfortably above the value of the noise floor of our amplifier.[6] We then calculated an attenuation ratio for each trace by assuming that the tunneling slope is 6 decades/nm. Next, a histogram of these attenuation ratios was constructed using the Freedman–Diaconis rule to determine the bin width, and finally we fit this histogram with a single unrestricted Gaussian (e.g. **Figure S2**). The peak of this Gaussian was taken as the attenuation ratio for all traces collected with the same sample. **Table S1** shows the number of tunneling traces used to calibrate each sample considered in this work and the resulting attenuation ratios.

**Table S1.** List of tunneling datasets considered in this work. One such tunneling dataset was collected for each MCBJ sample after depositing pure solvent (which quickly evaporates) but before depositing molecules. These tunneling datasets were used to determine an attenuation ratio for each sample, which was then applied to all subsequent datasets collected using that same sample.

| Dataset ID# | Sample # | Molecule Later Deposited on Sample | # of Traces | Attenuation Ratio/$10^{-4}$ | Solvent Used |
|---|---|---|---|---|---|
| 135 | 108-4 | C6-2SMe | 6460 | 1.01 | Hexanes |
| 127 | 111-4 | OPV2-2BT | 1537 | 1.02 | Dichloromethane |
| 130 | 108-5 | OPV2-2BT | 3847 | 1.52 | Dichloromethane |
| 49 | 097-2 | OPV3-BT-Br | 3469 | 2.01 | Dichloromethane |
| 82 | 106-1 | OPV3-BT-Br | 5777 | 1.96 | Dichloromethane |
| 85 | 098-4 | OPV3-BT-Cl | 2148 | 1.18 | Dichloromethane |
| 102 | 102-5 | OPV3-BT-Cl | 3881 | 2.30 | Dichloromethane |
| 105 | 101-4 | OPV3-BT-CN | 7664 | 1.58 | Dichloromethane |
| 112 | 101-3 | OPV3-BT-CN | 2084 | 1.55 | Dichloromethane |
| 117 | 114-2 | OPV3-BT-CN | 2206 | 1.42 | Dichloromethane |
| 120 | 103-2 | OPV3-BT-CN | 8009 | 1.98 | Dichloromethane |
| 30 | 098-2 | OPV3-BT-F | 4113 | 1.36 | Dichloromethane |
| 33 | 099-5 | OPV3-BT-F | 4894 | 1.47 | Dichloromethane |
| 94 | 099-1 | OPV3-BT-F | 4580 | 2.03 | Dichloromethane |
| 97 | 098-3 | OPV3-BT-F | 3454 | 1.72 | Dichloromethane |
| 1 | 113-2 | OPV3-BT-H | 3481 | 1.47 | Dichloromethane |
| 11 | 104-5 | OPV3-BT-H | 7122 | 1.17 | Dichloromethane |
| 54 | 104-4 | OPV3-BT-H | 3567 | 1.84 | Dichloromethane |
| 23 | 097-1 | OPV3-BT-MeO | 2523 | 1.27 | Dichloromethane |
| 27 | 111-2 | OPV3-BT-MeO | 2480 | 0.96 | Dichloromethane |
| 18 | 113-3 | OPV3-BT-NO2 | 6269 | 1.21 | Dichloromethane |
| 21 | 113-4 | OPV3-BT-NO2 | 3497 | 1.59 | Dichloromethane |

*S.3 Additional Design Criteria for Segment Clustering*

**S.3.1 Trace Starting and Ending Criteria.** In the BUS segmentation process, the first segment of each trace is forced to start at the first data point. It is thus important to use consistent starting criteria for every trace in a dataset to avoid any influence from confounding variables. For this work, we begin each trace the last time it passes below 2.5 $G_0$, to avoid issues with limited discrimination and accuracy of our amplifier at higher conductance values.[6] Modifications to these starting criteria do not meaningfully affect our results (see section S.5.3).

The ending criteria for each trace are similarly important. We first exclude any conductances below the noise floor of our amplifier[6] (typically $10^{-6}$ $G_0$, but slightly higher in a few datasets; see section S.4 for details). Additionally, in cases where a trace drops below the noise floor but then later returns to a higher conductance, we end the trace the *first* time it drops below this level. This is necessary to avoid large empty gaps in traces, since BUS is not designed to work in such cases.

**S.3.2 Parameter Standardization.** Standardizing the distribution of a variable typically involves dividing by the sample standard deviation. However, because the standard deviation is sensitive to outliers, this can skew the standardization process. In contrast, the range of the middle 80% of values in a dataset is quite insensitive to outliers, so we use this measure of spread to standardize the first three segment parameters ($X_0$, $Y_0$, and $\text{Log}(L)$). Because the $\theta$ and $R^2$ parameters have limited possible ranges—(-90° to 90°) and (0 to 1), respectively—we standardize them by dividing by 80% of those full possible ranges. This has the advantage of making the standardization process less dependent on a particular dataset.

Finally, $\theta$ is only calculated *after* the inter-electrode distance and log(conductance) dimensions have already been standardized. This is necessary to ensure that $\theta$ is fully independent of the units on the x- and y-axes.

**S.3.3 Assignment of SOPTICS Parameter Values.** The $c_L$ and $c_P$ SOPTICS parameters control how many random projections are performed, with larger values leading to a more stable and accurate approximation of the original OPTICS algorithm. The creators of SOPTICS found that $c_L = c_P = 20$ produced excellent results;[7] out of an abundance of caution, we use a higher value of $c_L = c_P = 30$ (see section S.5.1 for details).

The *minSize* parameter helps control how the random projections are sampled to find nearby points.[7] Because SOPTICS is extremely insensitive to the value of *minSize* over a large range (see section S.5.2), we fix its value at 120.

**S.3.4 Length-Weighting of Segments.** Because OPTICS/SOPTICS is a density-based clustering algorithm, the density of segment parameters in 5-dimensional space ultimately determines how segments are clustered, with the densest regions forming the "cores" of key clusters. However, because segments are drawn from traces of roughly the same length, there will almost always be many more short segments than long ones. Perversely, this leads to a *lower* density of long segments, even though they represent many *more* data points from the original traces, making it difficult to form clusters of long segments. To remedy this issue, in the density calculations we apply a weighting factor to each segment that is proportional to its length. This ensures that the density of segments in parameter space corresponds to the density of raw data points rather than the number of segments themselves. In practice, this weighting is accomplished by duplicating each segment in proportion to its length before clustering. This step introduces another parameter, *len_per_dup*, that controls how many times a segment of a given length is duplicated. This parameter also serves as the minimum segment length, as we exclude segments that are not long enough for even a single duplicate. We set *len_per_dup* to 0.05 nm (*e.g.*, segments between 0.20 and 0.24 nm long will have 4 total copies) to ensure that all segments down to the length of a single bond will be included. We also note that the effects of changing *len_per_dup* are correlated with the effects of changing *minPts*, the parameter that defines how density is estimated (see section S.5.4). Therefore, because we use 12 different values of *minPts*, we are already capturing much of the possible variation from using different values of *len_per_dup*. Segment duplication is performed after the parameterization step.

*S.4 Dataset Collection and Construction*

Pausing a sample to re-deposit molecular solution often leads to a discrete, qualitative change in trace behavior—e.g., the fraction of traces displaying a molecular plateau (the "molecular yield") may significantly increase or decrease after re-deposition, or the gold electrodes may undergo rearrangement, as evidenced by a significant change in where bridge rupture occurs on the absolute push rod movement scale. Such changes may also occur when depositing pure solvent on a junction already containing molecules, or when "starting a new trial" by fully relaxing the push rod and junction, followed by restarting trace-collection. We therefore treat the traces collected during each deposition/trial combination for a given MCBJ sample as a separate dataset. In the context of clustering, splitting each sample into multiple datasets in this way is the conservative approach; if instead we clustered the traces from each sample as one big dataset, we would be much more likely to find "consistent" features because the algorithm might only identify the regions where multiple disparate features all overlap. Splitting datasets at the natural points where qualitative changes tend to occur challenges Segment Clustering by providing the most opportunities for it to be confounded by changes in the "background".

For this work, we did not consider datasets from samples which showed strong signs of contamination in their initial "pure tunneling" sections. We also excluded molecular datasets in which no molecular feature was apparent, insufficient traces were collected (significantly less than 1000), or obvious noise features were present. For the OPV3-2BT-X family, this left us with 43 different molecular datasets, each corresponding to an entire deposition/trial block of traces (**Table S2**). We observed no apparent correlation between the number of depositions or trials and junction conductance. In nearly all of these datasets, the noise floor was set to $10^{-6}$ $G_0$, the nominal bottom end of the range for our amplifier.[6] However, due to differences in calibration, a few samples displayed higher noise levels, requiring us to manually set a higher noise floor to prevent physically meaningless data from affecting clustering results (see **Table S2**).

**Table S2.** List of all OPV3-2BT-X datasets considered in this work. Each dataset corresponds to a full deposition/trial block of traces. All molecular solutions were 1 μM. The top-to-bottom order of datasets in this table corresponds with the left-to-right order of points in Figure 7.

| Dataset ID# | Sample # | Trial # | Deposition # | # of Traces | Molecule Name | Noise Floor ($G_0$) |
|---|---|---|---|---|---|---|
| 2 | 113-2 | 1 | 1 | 5424 | OPV3-BT-H | 1.0E-06 |
| 3 | 113-2 | 1 | 2 | 9446 | OPV3-BT-H | 1.0E-06 |
| 12 | 104-5 | 1 | 1 | 3545 | OPV3-BT-H | 1.0E-06 |
| 13 | 104-5 | 1 | 2 | 4550 | OPV3-BT-H | 1.0E-06 |
| 14 | 104-5 | 2 | 2 | 2997 | OPV3-BT-H | 1.0E-06 |
| 15 | 104-5 | 3 | 2 | 6280 | OPV3-BT-H | 1.0E-06 |
| 16 | 104-5 | 4 | 2 | 5062 | OPV3-BT-H | 1.0E-06 |
| 58* | 104-4 | 2 | 3 | 4113 | OPV3-BT-H | 1.0E-06 |
| 59 | 104-4 | 3 | 3 | 6294 | OPV3-BT-H | 1.0E-06 |
| 25 | 097-1 | 2 | 1 | 4065 | OPV3-BT-MeO | 1.0E-06 |
| 26 | 097-1 | 2 | 2 | 3137 | OPV3-BT-MeO | 1.0E-06 |
| 28 | 111-2 | 2 | 1 | 4051 | OPV3-BT-MeO | 1.0E-06 |
| 29 | 111-2 | 2 | 2 | 6214 | OPV3-BT-MeO | 1.0E-06 |
| 31 | 098-2 | 1 | 1 | 5182 | OPV3-BT-F | 1.0E-06 |
| 95 | 099-1 | 1 | 1 | 7695 | OPV3-BT-F | 1.0E-06 |
| 96 | 099-1 | 1 | 2 | 2147 | OPV3-BT-F | 1.0E-06 |
| 34 | 099-5 | 1 | 1 | 7922 | OPV3-BT-F | 1.0E-06 |
| 35 | 099-5 | 1 | 2 | 18568 | OPV3-BT-F | 1.0E-06 |
| 37 | 099-5 | 2 | 3 | 3941 | OPV3-BT-F | 1.0E-06 |
| 98 | 098-3 | 1 | 2 | 8661 | OPV3-BT-F | 1.0E-06 |
| 99 | 098-3 | 1 | 4 | 8753 | OPV3-BT-F | 1.0E-06 |
| 101 | 098-3 | 2 | 5 | 4120 | OPV3-BT-F | 1.0E-06 |
| 86 | 098-4 | 2 | 1 | 2940 | OPV3-BT-Cl | 1.0E-06 |
| 88 | 098-4 | 3 | 2 | 7670 | OPV3-BT-Cl | 1.0E-06 |
| 103 | 102-5 | 1 | 1 | 6394 | OPV3-BT-Cl | 1.0E-06 |
| 104 | 102-5 | 1 | 2 | 7841 | OPV3-BT-Cl | 1.0E-06 |
| 50 | 097-2 | 1 | 1 | 8603 | OPV3-BT-Br | 1.0E-06 |
| 51 | 097-2 | 1 | 2 | 10529 | OPV3-BT-Br | 1.0E-06 |
| 83 | 106-1 | 1 | 1 | 9572 | OPV3-BT-Br | 1.0E-06 |
| 84 | 106-1 | 1 | 2 | 15707 | OPV3-BT-Br | 1.0E-06 |
| 19 | 113-3 | 1 | 1 | 7310 | OPV3-BT-NO2 | 1.0E-06 |
| 20 | 113-3 | 1 | 2 | 8083 | OPV3-BT-NO2 | 1.0E-06 |
| 22 | 113-4 | 1 | 1 | 7799 | OPV3-BT-NO2 | 1.0E-06 |
| 107 | 101-4 | 2 | 2 | 6679 | OPV3-BT-CN | 5.5E-06 |
| 108 | 101-4 | 2 | 3 | 7449 | OPV3-BT-CN | 5.5E-06 |
| 109 | 101-4 | 2 | 4 | 2309 | OPV3-BT-CN | 5.5E-06 |
| 114* | 101-3 | 1 | 2 | 2772 | OPV3-BT-CN | 1.0E-05 |
| 116 | 101-3 | 2 | 3 | 5477 | OPV3-BT-CN | 1.0E-05 |
| 118 | 114-2 | 1 | 3 | 4280 | OPV3-BT-CN | 3.0E-06 |
| 121 | 103-2 | 1 | 1 | 10259 | OPV3-BT-CN | 1.0E-06 |
| 123 | 103-2 | 2 | 2 | 3175 | OPV3-BT-CN | 1.0E-06 |
| 125 | 103-2 | 3 | 3 | 2783 | OPV3-BT-CN | 1.0E-06 |
| 126** | 103-2 | 3 | 3 | 6548 | OPV3-BT-CN | 1.0E-06 |

*Dataset not included in analysis; see section S.7 for details.
**Pure dichloromethane was deposited between datasets #125 and #126; hence they are treated as distinct datasets even though they have the same trial number and number of molecular depositions.

For this work, we also considered five OPV2-2BT datasets and two C6-2SMe datasets (**Table S3**). In two of these cases, the dataset consisted of a subset of consecutive traces from a deposition/trial block in order to exclude clear noise features (see **Table S3**). We then constructed eight different 1:1 synthetic mixtures of these OPV2-2BT and C6-2SMe datasets by combining different sets of traces from different datasets. Because the OPV2-2BT datasets contained more traces, for each mixture we used *all* of the traces from one of the C6-2SMe datasets and then added an equivalent number of consecutive traces from a subset of one of the OPV2-2BT datasets (see **Table S4** for details).

**Table S3.** List of all OPV2-2BT and C6-2SMe datasets considered in this work. "Subset" refers to datasets corresponding to a consecutive subset of traces from an entire deposition/trial block, taken to exclude clear noise features. All noise floors are $10^{-6}$ $G_0$.

| Dataset ID# | Sample # | Trial # | Deposition # | Subset | # of Traces | Molecule Name | Solution Concentration (µM) | Solvent Used |
|---|---|---|---|---|---|---|---|---|
| 128 | 111-4 | 1 | 1 | No | 3234 | OPV2-2BT | 1 | Dichloromethane |
| 129 | 111-4 | 1 | 2 | No | 2680 | OPV2-2BT | 1 | Dichloromethane |
| 132 | 108-5 | 1 | 2 | Yes | 2400 | OPV2-2BT | 1 | Dichloromethane |
| 133 | 108-5 | 3 | 5 | No | 6562 | OPV2-2BT | 1 | Dichloromethane |
| 134 | 108-5 | 4 | 1* | No | 5807 | OPV2-2BT | 10 | Dichloromethane |
| 136 | 108-4 | 1 | 2 | Yes | 1315 | C6-2SMe | 10 | Hexanes |
| 137 | 108-4 | 2 | 2 | No | 1065 | C6-2SMe | 10 | Hexanes |

*1st deposition of a 10 µM solution, but 6th depositon overall (first 5 depositions were each with a 1 µM solution).

Since each sample has a slightly different attenuation ratio, the density of data points on the inter-electrode distance scale is also different for each sample. This is an issue for constructing synthetic mixture datasets because it would cause the denser dataset to have extra weight in what is supposed to be a 1:1 mixture. We therefore used linear interpolation to resample all OPV2-2BT and C6-2SMe traces at a rate of one data point per $4\times10^{-4}$ nm of inter-electrode distance. This resampling was performed before clustering the pure datasets and before the construction and clustering of the synthetic mixture datasets.

**Table S4.** List of the eight different 1:1 OPV2-2BT:C6-2SMe synthetic mixture datasets created for this work, along with details of their construction. Dataset ID #s refer to **Table S3**. Mixture #1 is the dataset used for Figure 8g-i.

| Mixture # | Total # of Traces | Dataset ID for OPV2-2BT Traces | Dataset ID for 2,9-dithiadecane | Traces Used from OPV2-2BT Dataset |
|---|---|---|---|---|
| 1 | 2630 | 134 | 136 | 1-1315 |
| 2 | 2130 | 134 | 137 | 1-1065 |
| 3 | 2630 | 133 | 136 | 1-1315 |
| 4 | 2630 | 128 | 136 | 1-1315 |
| 5 | 2630 | 132 | 136 | 1-1315 |
| 6 | 2130 | 132 | 137 | 1-1065 |
| 7 | 2630 | 129 | 136 | 1-1315 |
| 8 | 2630 | 134 | 136 | 1500-2814 |

*S.5 Robustness of OPV3-2BT-X Results to Clustering Parameters*

**S.5.1 Robustness to Random Seed.** The SOPTICS algorithm employs random projections in order to achieve its improved clustering times, and even regular OPTICS, when properly implemented, uses a random choice for the first point in the cluster order. If the clustering structure extracted by these algorithms is truly inherent to the dataset, then the clustering results should not be meaningfully affected by using a different set of random numbers. To confirm that this is the case for our OPV3-2BT-X results, we re-clustered one of our datasets (ID# 3 in **Table S2**) using ten different random seeds for MatLab's pseudo-random number generator. This is also a good way to evaluate our choice for the parameters $c_L$ and $c_P$; because these parameters control how many different random projections are used by SOPTICS, we know that their values are suitably large when the clustering outputs for different random seeds all converge to give the same results. We therefore repeated this random seed testing for three different values of $c_L = c_P$. For this testing we fixed the value of *minPts* at 85.

We used two different methods to evaluate the similarity of these different clustering results. First, we simply compared the peak conductance value for the main plateau cluster in this dataset, as this peak conductance is what we are ultimately interested in for our analysis of the OPV3-2BT-X family. Second, we used a similarity index developed by Rand to compare the entire clustering *solutions* that each main plateau cluster belongs to. The Rand similarity index is a pairwise comparison that ranges from 0 to 1, with 1 meaning that every data point was assigned to the same cluster in both clustering solutions and 0 meaning that every data point was assigned to a different cluster in one solution vs. the other.[8] Because this method compares the overall clustering structure instead of just the peak value of a single cluster, it provides a more stringent test of the similarity of different clustering results.

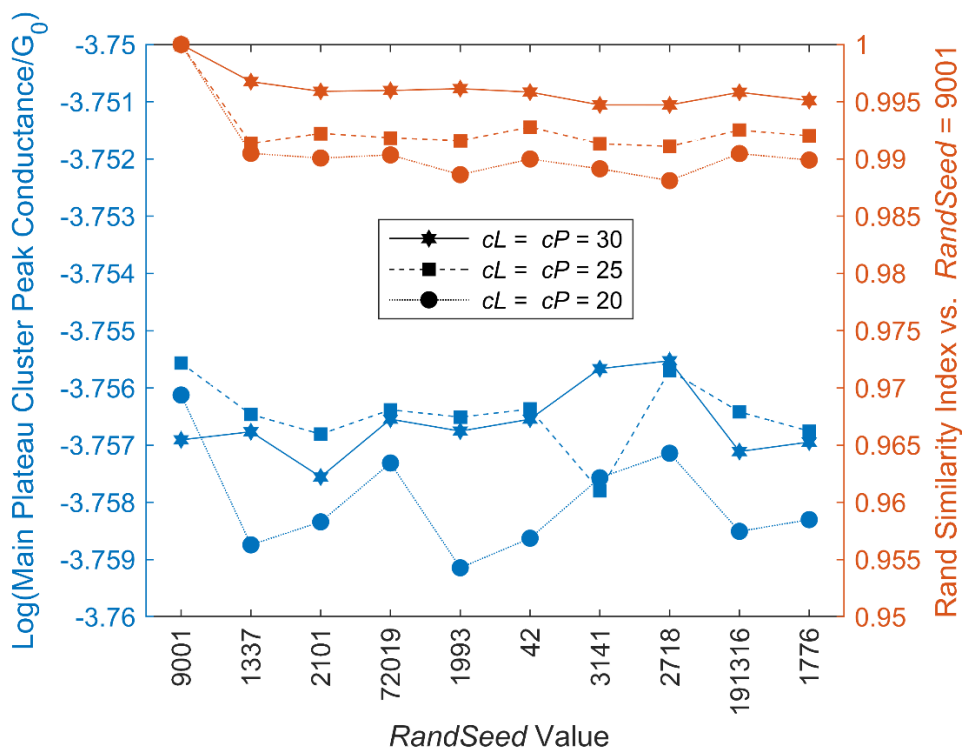

**Figure S3.** Comparison of fitted peak conductance values for the main plateau cluster for a single OPV3-2BT-H dataset clustered using 10 different random seeds and three different values for the parameters *cL* = *cP*, with the *minPts* parameter fixed at 85 (left axis). For the right axis, the clustering *solution* which contained the main plateau cluster for each of the 30 clustering outputs was first identified. Each of these solutions was then compared to the solution for a random seed of 9001 using the Rand similarity index. These results demonstrate both that SOPTICS is not affected meaningfully by random seed choice and that *cL* = *cP* is set to a sufficiently large value.

The results of these evaluation methods for our random seed testing are summarized in

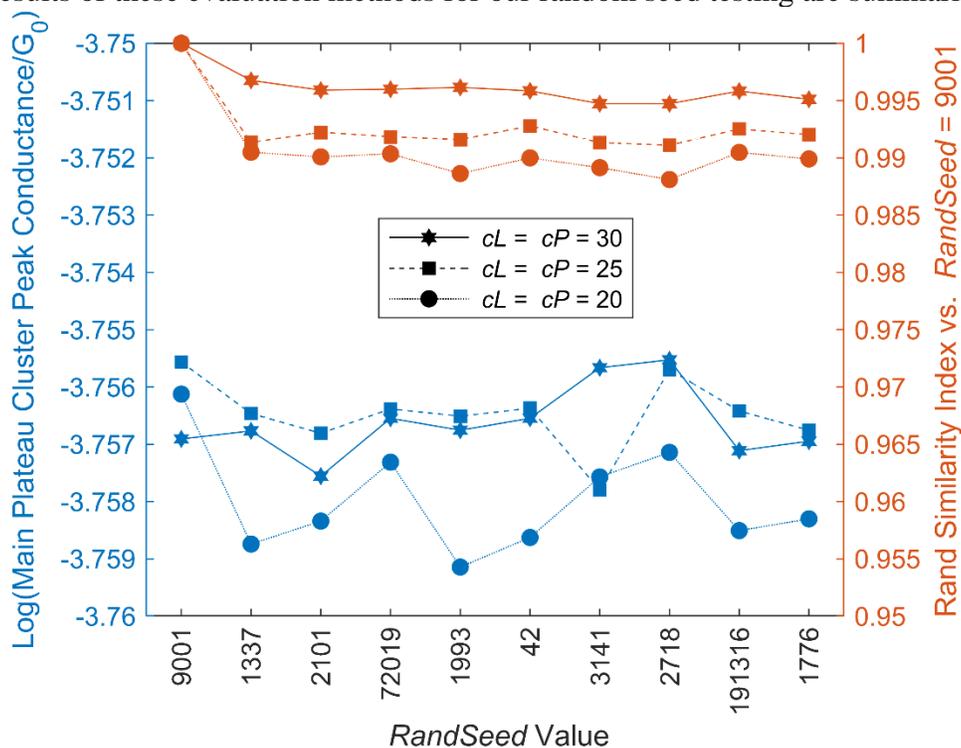

**Figure S3**. Even for $c_L = c_P = 20$, changing the random seed has essentially no effect, with the conductance peak varying by less than 0.003 decades and the Rand Similarity Index always greater than 0.985. Our decision to use $c_L = c_P = 30$, where the convergence is even tighter, is thus clearly a very safe choice.

In addition, these results demonstrate that in our implementation SOPTICS is essentially unaffected by the set of random numbers used, and is thus behaving properly. For the clustering results discussed in the main body of the paper and for all subsequent testing, we therefore used the last digits of the system time to generate a different random seed for each clustering run.

**S.5.2 Robustness to *minSize*.** To ensure that our OPV3-2BT-X results are not dependent upon our choice for the *minSize* parameter, we re-clustered another dataset (ID# 25 in **Table S2**) using 17 different values of *minSize*. We again fixed the value of *minPts* at 85 for this testing.

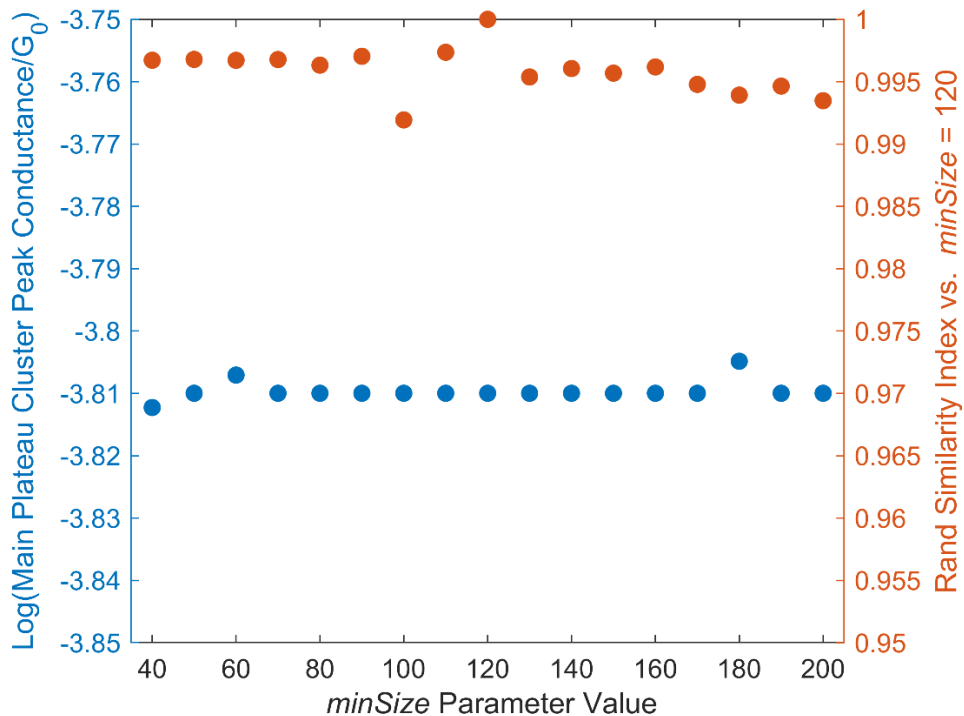

**Figure S4.** Comparison of fitted peak conductance values for the main plateau cluster for a single OPV3-2BT-MeO dataset clustered using 17 different values for the parameter *minSize*, with the *minPts* parameter fixed at 85 (left axis). For the right axis, the clustering *solution* which contained the main plateau cluster for each of the 17 clustering outputs was first identified. Each of these solutions was then compared to the solution for *minSize* = 120 using the Rand similarity index. These results demonstrate that the exact value of *minSize* is not very important for the behavior of SOPTICS, and so it is safe to use a single fixed value for this parameter.

S14

We used the same two evaluation methods (main plateau cluster peak conductance and Rand similarity index) described in section S.5.1 to compare these different clustering results. As shown in

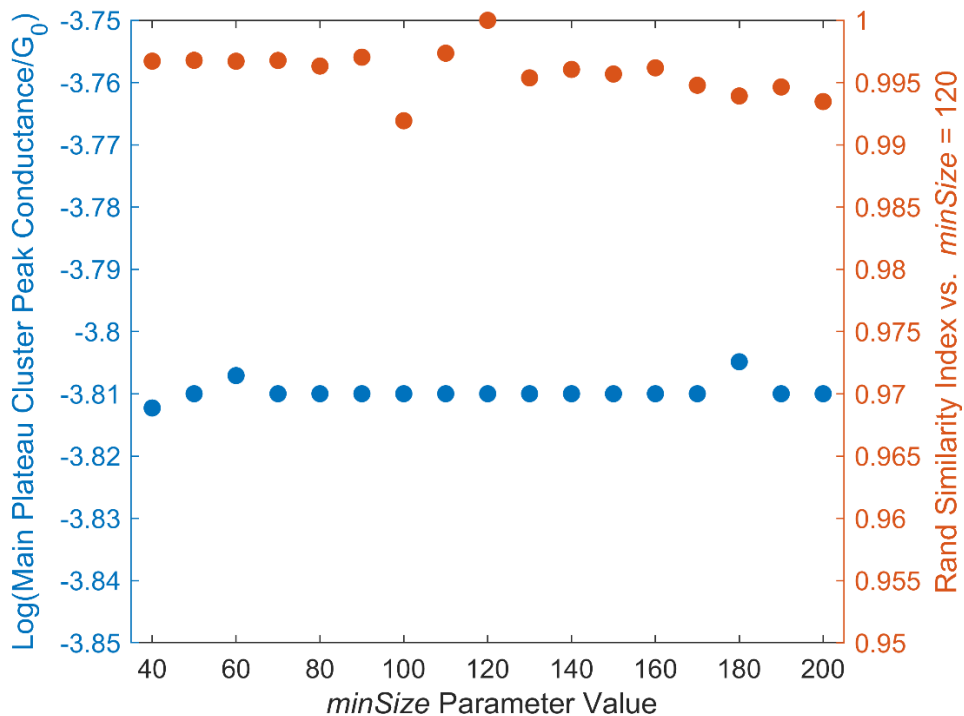

**Figure S4**, the clustering output is extremely insensitive to the choice of *minSize* over quite a large range. This justifies our choice to fix the value of *minSize* at 120.

**S.5.3 Robustness to Trace Starting Criteria.** As described in section S.3.1, to ensure consistent starting criteria before the segmentation step, we begin each trace the last time it passed below a conductance of 2.5 $G_0$. To check that our OPV3-2BT-X results do not depend on this choice, we re-clustered another of our datasets (ID# 19 in **Table S2**) using six different values for this "*TopChop*" conductance value.

Because changing the *TopChop* affects the segmentation step, these different clustering outputs do not contain the exact same objects for clustering, and so cannot be compared using the Rand similarity index.

However, comparing the peak conductance of the main plateau cluster for each of these results (

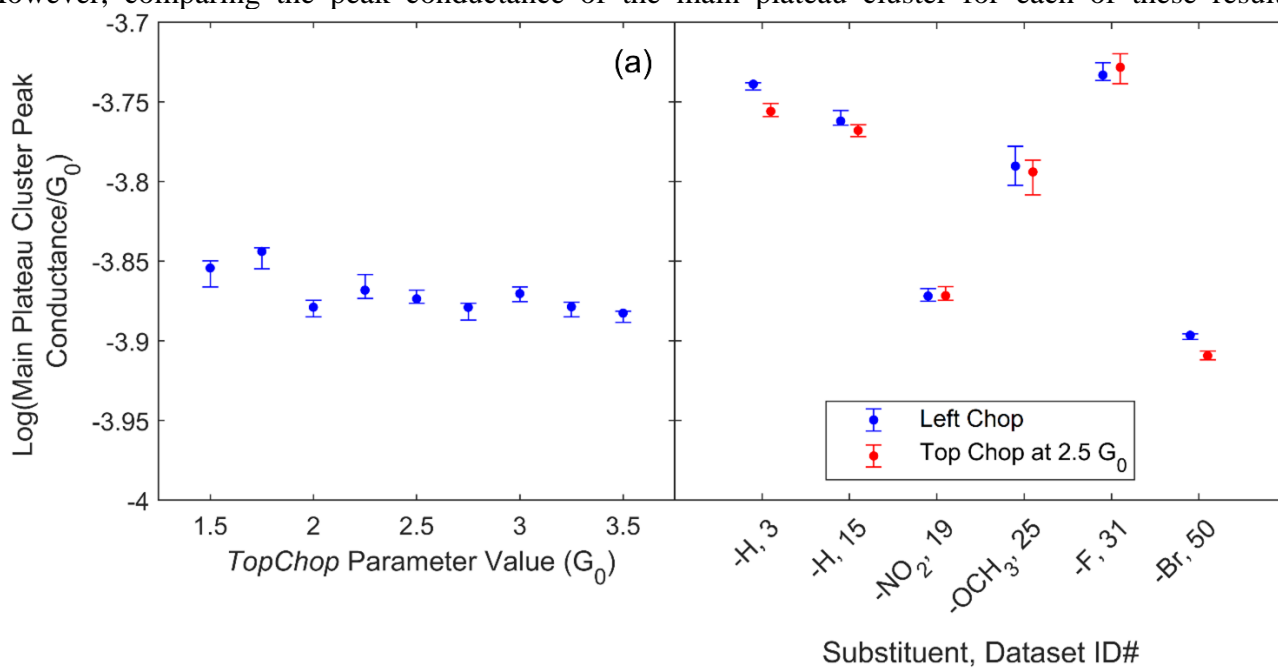

**Figure S5a**) shows that the choice of the *TopChop* value does not meaningfully impact our results.

As an additional test, we also considered a different type of starting criteria: instead of a "*TopChop*", a "*LeftChop*" in which we begin each trace at zero inter-electrode distance. Comparing the results for six of our datasets for these two different chop methods (

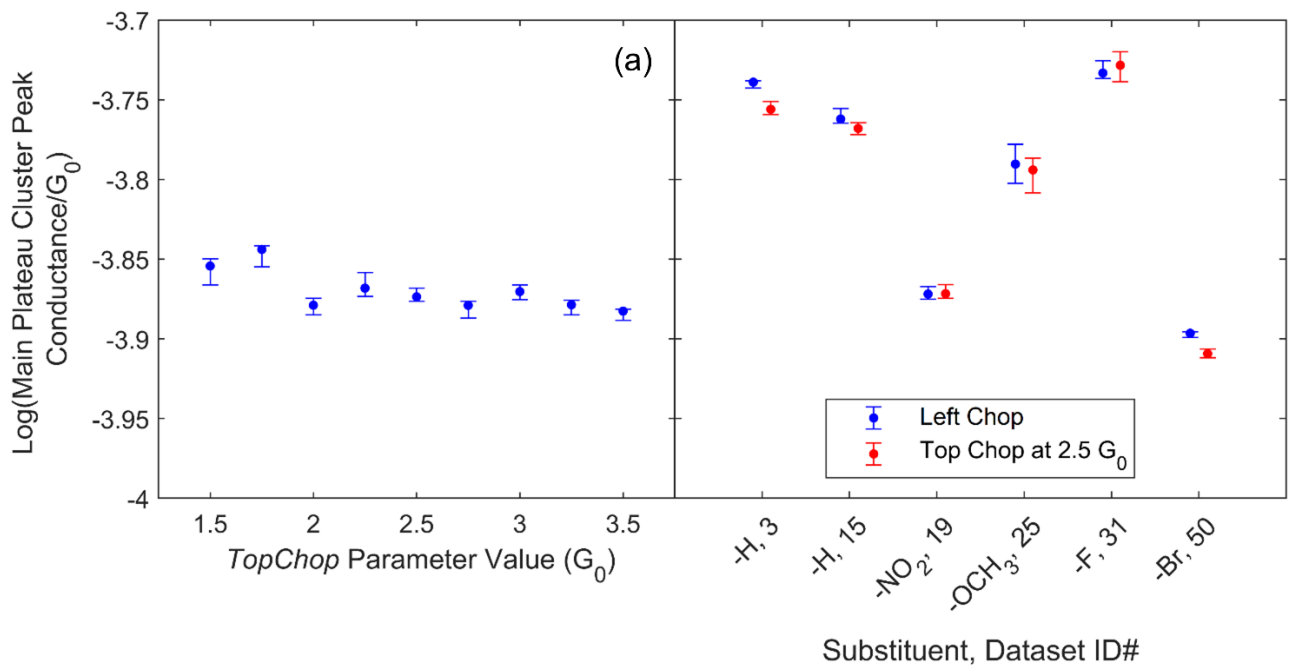

**Figure S5b**) again confirms that our OPV3-2BT-X conclusions are not dependent upon our choice of starting criteria. We note that this left chop at zero significantly improves clustering time by reducing the number of data points, and so may be preferred in some situations.

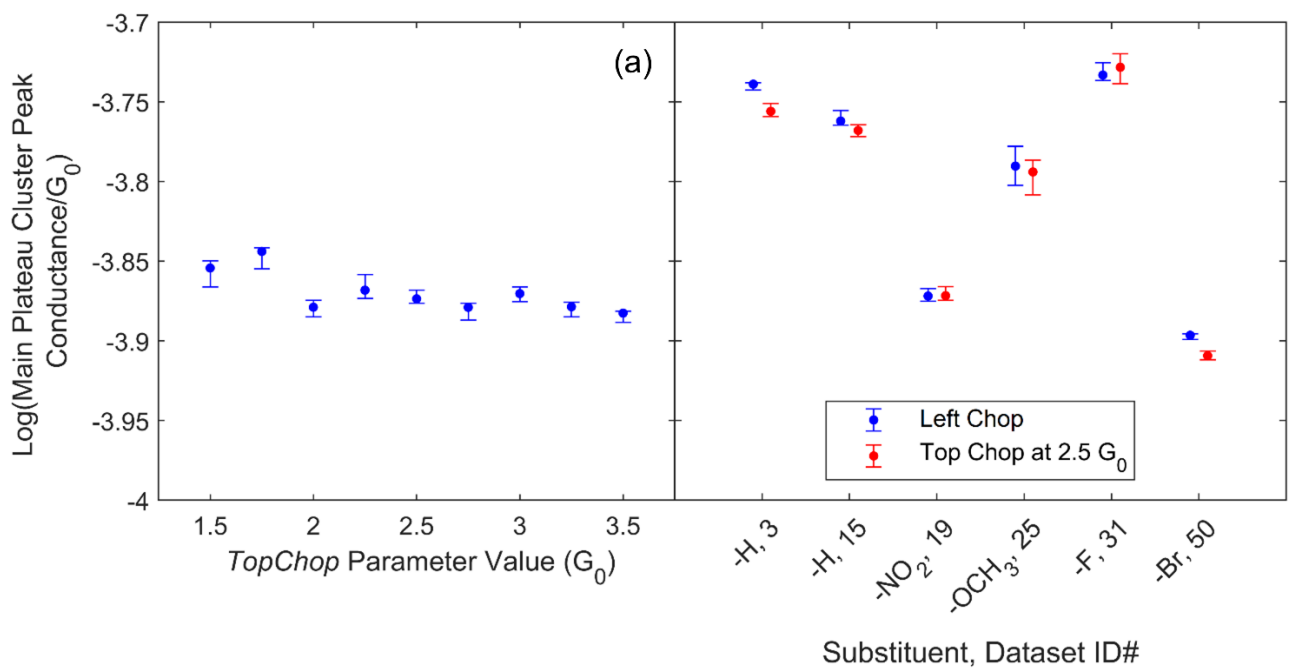

**Figure S5.** Demonstration of the insensitivity of OPV3-2BT-X clustering results to trace starting criteria. (a) Peak conductance values for the main plateau cluster of the same OPV3-2BT-NO$_2$ dataset clustered using 9 different "*TopChop*" values (only the portion of each trace after the last time its conductance passes below *TopChop* is included for clustering). (b) Comparison of the peak conductance values for the main plateau clusters for six different OPV3-2BT-X datasets (dataset ID#s refer to **Table S2**) clustered using a *TopChop* of 2.5 G$_0$ (red) or a "*LeftChop*" (blue), in which only the portion of each trace after zero inter-electrode distance is included for clustering.

**S.5.4 Robustness to *len_per_dup* and Correlation with *minPts*.** As described in section S.3.4, the parameter *len_per_dup* controls how often each segment is duplicated in proportion to its length (and also sets the minimum segment length). Decreasing *len_per_dup* increases the density of data points in all regions, and is thus expected to have a similar effect to decreasing the value of *minPts*. To confirm this, we re-clustered one of our datasets (ID# 2 in **Table S2**) at a variety of combinations of *minPts* and *len_per_dup* parameter values. The clustering solutions containing the main plateau cluster were then compared using the Rand similarity index as well as the peak plateau conductance (

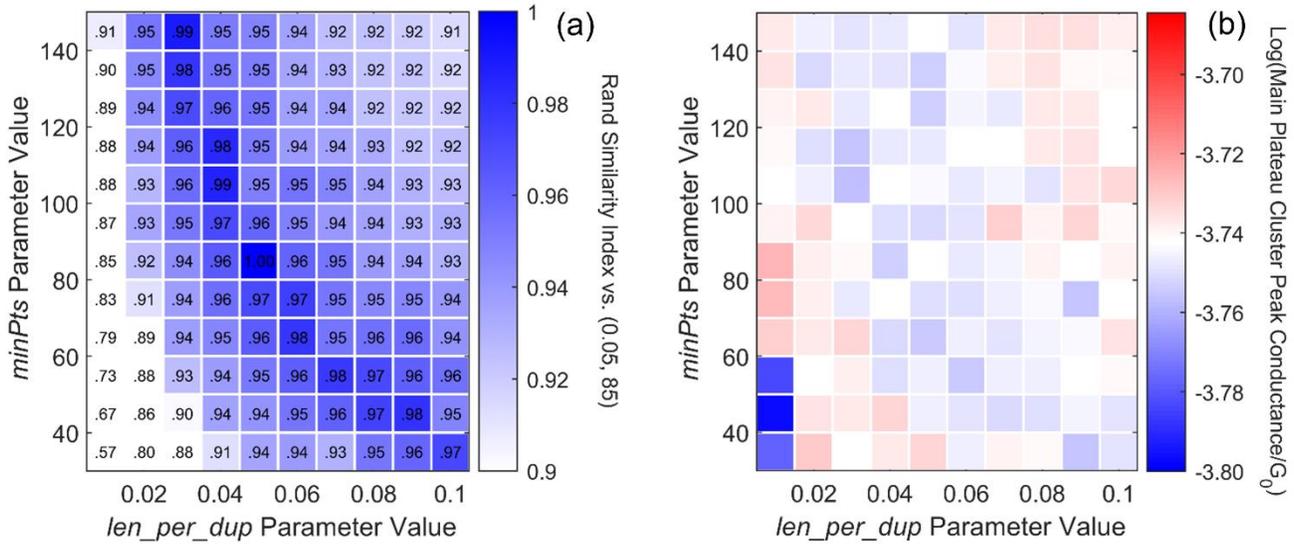

**Figure S6a,b**), as described in section S.5.1. Because *len_per_dup* also controls the minimum segment length, clustering runs with larger *len_per_dup* values used slightly fewer segments for clustering. Therefore, for each pairwise comparison only those segments present in *both* clustering results were considered when computing the Rand similarity index.

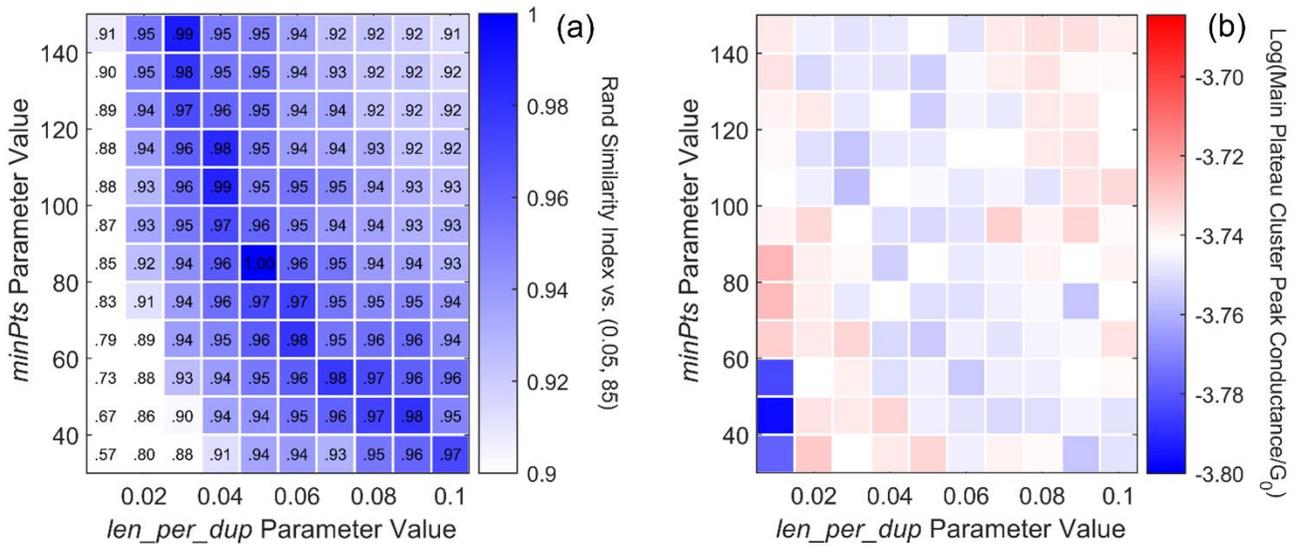

**Figure S6.** Comparison of outputs for a single OPV3-2BT-H dataset clustered using 120 different combinations of the *minPts* and *len_per_dup* parameters. (a) Rand similarity index for the clustering solution from each output which contained the main plateau cluster, compared to the chosen solution for the *minPts* = 85 and *len_per_dup* = 0.05 nm output. The fact that most of the index values are close to one shows that the clustering is relatively insensitive to these two parameters, and the northwest-to-southeast "ridge" demonstrates that they are positively correlated with each other. (b) Fitted peak

conductance values for the main plateau cluster for each output, demonstrating that this measurement is quite insensitive to both parameters.

The high Rand similarity indices (

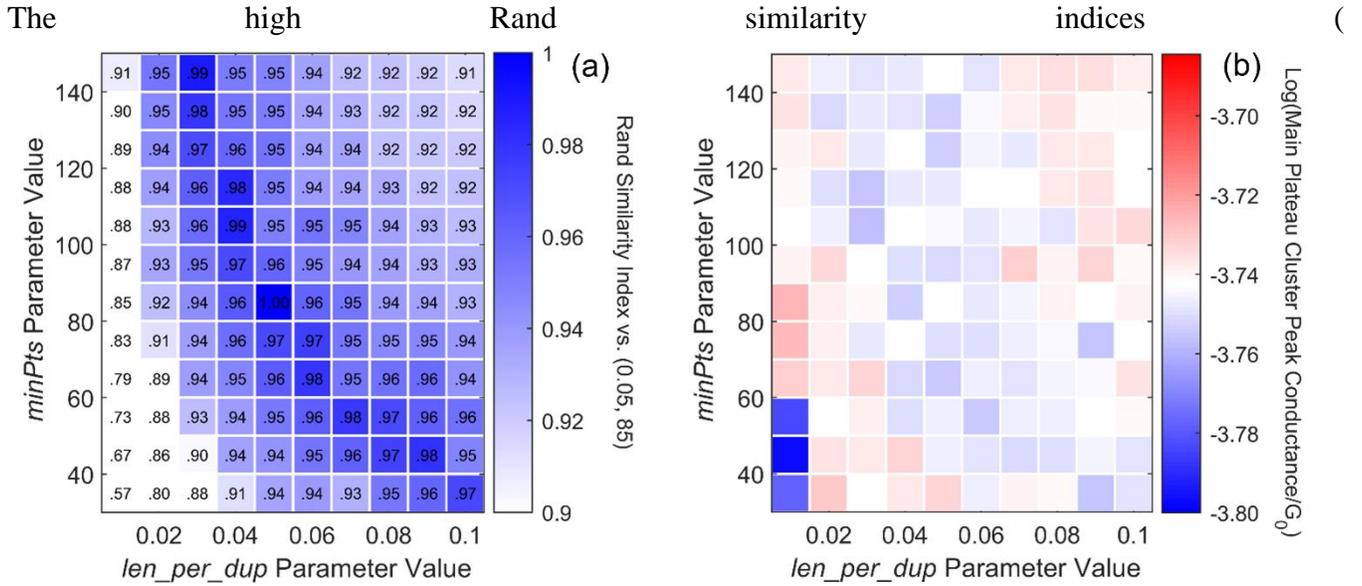

Figure S6a) and similar peak conductance values (

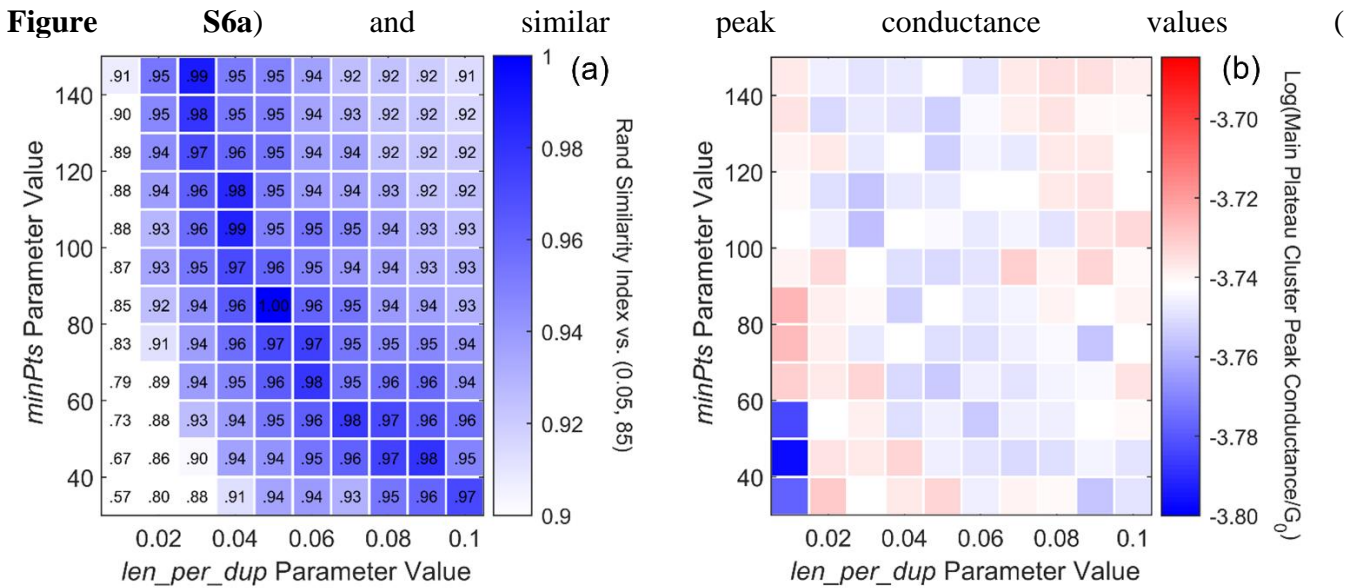

**Figure S6b**) that are found across a wide range of *len_per_dup* values indicate that clustering results are quite robust to changes in this parameter. More importantly, however,

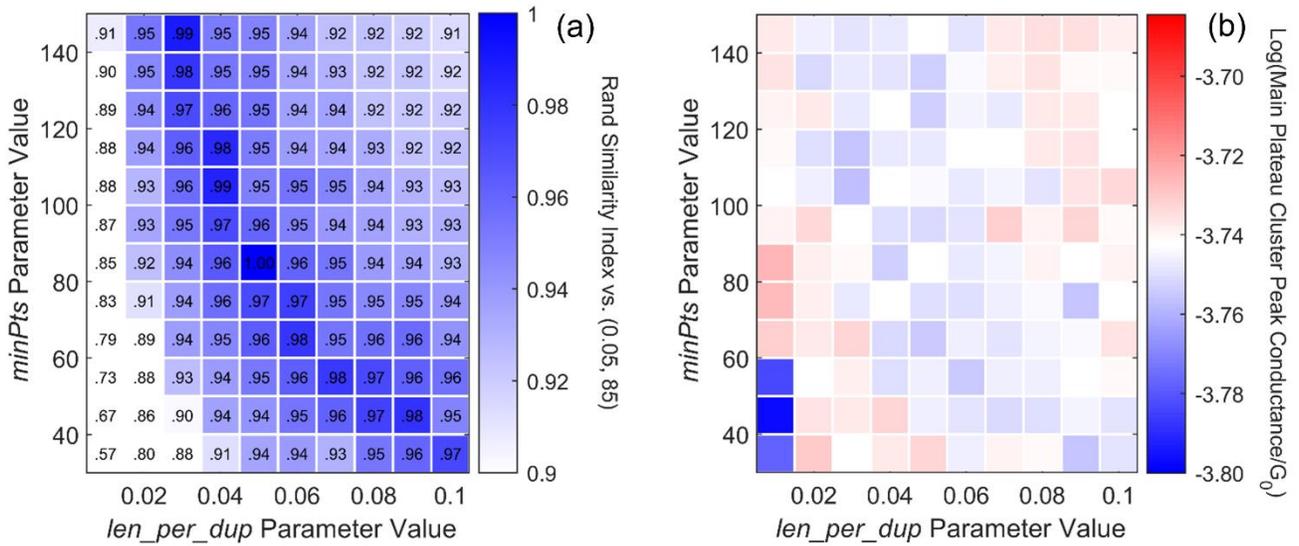

**Figure S6a** demonstrates that there is indeed a strong correlation between the effects of changing the *len_per_dup* and *minPts* parameters, as expected. This helps justify our decision to fix the value of -*len_per_dup*, because it means that by using multiple values of *minPts* we are already capturing much of the variation that would be caused by changes to *len_per_dup*.

**S.5.5 Robustness to Settings of Iterative L-Method.** One of the advantages to using the Iterative L-Method as a stopping criterion for Bottom-Up Segmentation is that it is described as being parameter-free. However, the algorithm does rely on a value, *minimum_cutoff_size*, which the authors argue can be considered a constant instead of a parameter because a value of 20 yields good results in a wide variety of contexts.[9] Out of an abundance of caution, we also tried re-clustering a handful of our datasets using a smaller (16) or larger (24) value of *minimum_cutoff_size*.

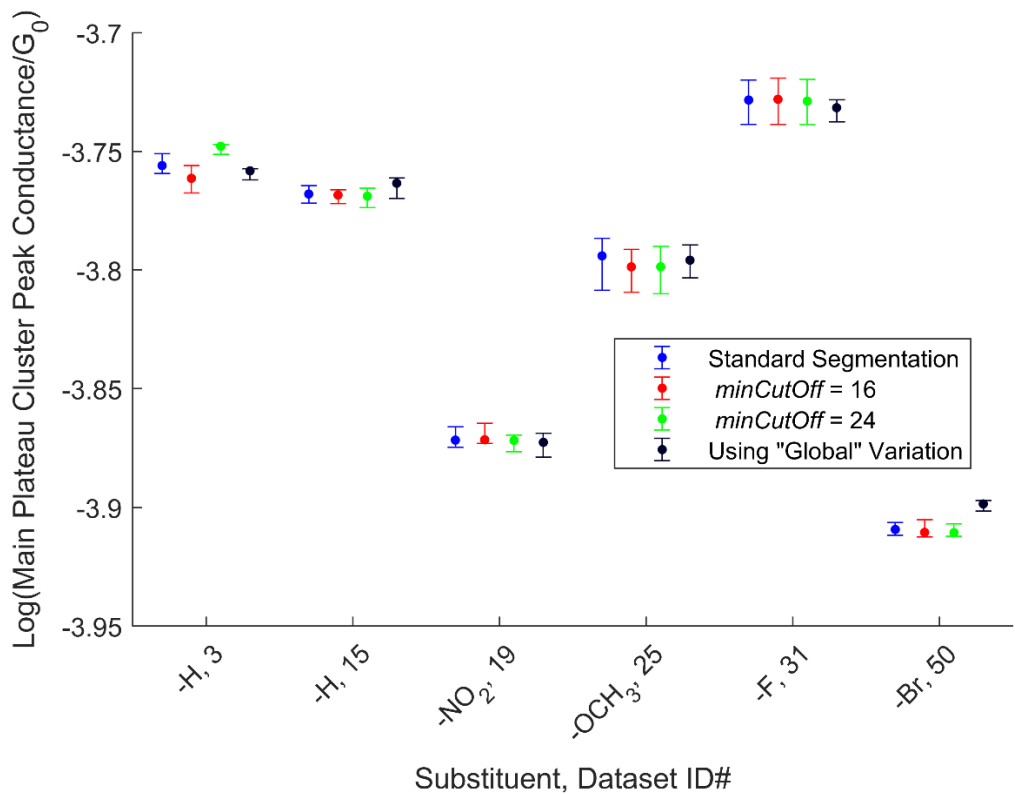

**Figure S7.** Comparison of the peak conductance values for the main plateau clusters for six different OPV3-2BT-X datasets (dataset ID#s refer to **Table S2**) clustered after using the "standard" segmentation procedure (blue); after segmentation with the *minimum_cutoff_size* value set to 16 (red) or 24 (green) instead of its standard value of 20; and after using the "Global" instead of the "Greedy" Iterative L-Method as stopping criteria for segmentation (black). These results demonstrate that slight variations in how the segmentation algorithm is implemented do not meaningfully affect our OPV3-2BT-X results.

Additionally, the authors actually present two slight variations of the Iterative L-Method: "Global" and "Greedy". As mentioned above, we use the "Greedy" Iterative L-Method because it was generally found to produce superior results.[9] However, again out of an abundance of caution, we also tried re-clustering these same datasets using the "global" Iterative L-Method instead. As shown in

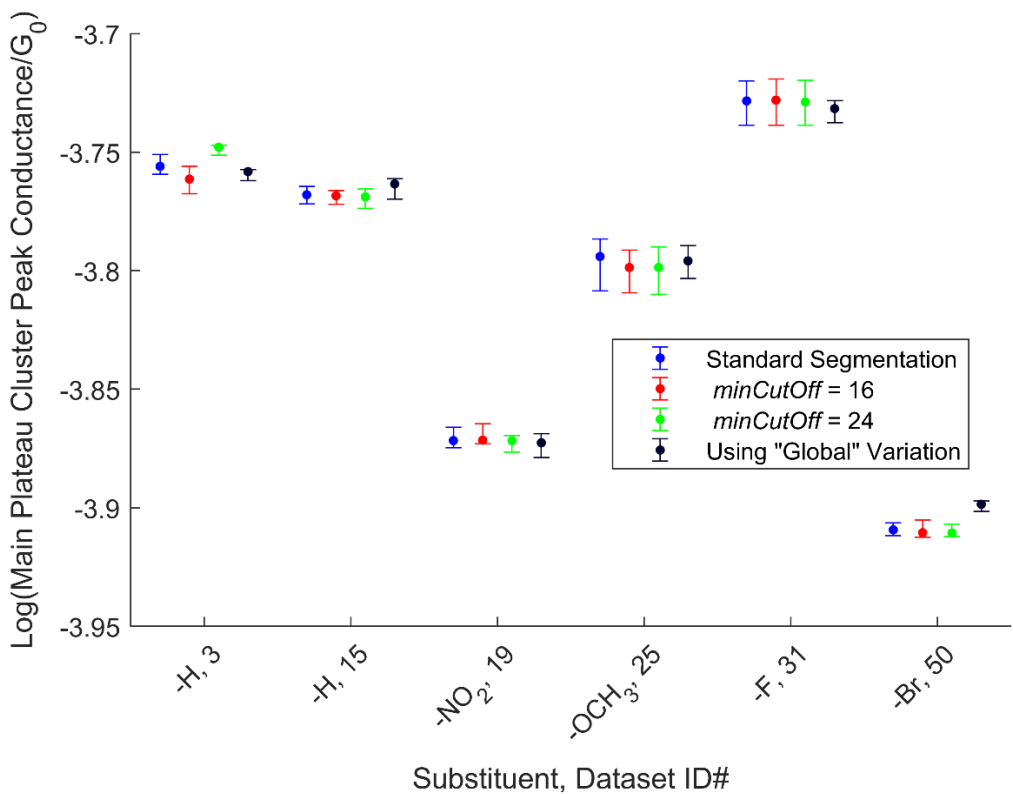

**Figure S7**, neither the changes to *minimum_cutoff_size* nor the switch from "Greedy" to "Global" meaningfully affect our results for the OPV3-2BT-X molecules.

*S.6 Selecting Clusters from Multiple Cluster Outputs for the Same Dataset*

As discussed in the main text, each dataset in this work was re-clustered twelve times using different values of the parameter *minPts* in order to account for uncertainty in the "optimal" setting for this parameter. For the figures in this work, we calculated and show each clustering output for *minPts* = 85 (roughly in the center of the 12 different *minPts* values).

After selecting a particular full-valley cluster of interest in the *minPts* = 85 output of a given dataset (e.g. the main plateau cluster for each OPV3-2BT-X dataset), we employed an automated algorithm to identify the analogous full-valley cluster in each of the other eleven clustering outputs for that same dataset. This algorithm first calculates the median value of each normalized segment parameter for the manually chosen cluster as well as for every full-valley cluster in the other eleven outputs. It then selects the single full-valley cluster from each of those outputs with the smallest Euclidean distance between its "median centroid" and that of the manually chosen cluster. The clusters identified with this automated algorithm matched the unambiguous assignments that would have been made by eye.

When the distributions for chosen clusters were fit to determine peak conductance values, the clusters from the twelve different outputs for each dataset were fit independently to obtain twelve different peak values. To represent the peak conductance of a single dataset (specifically, in Figure 7,

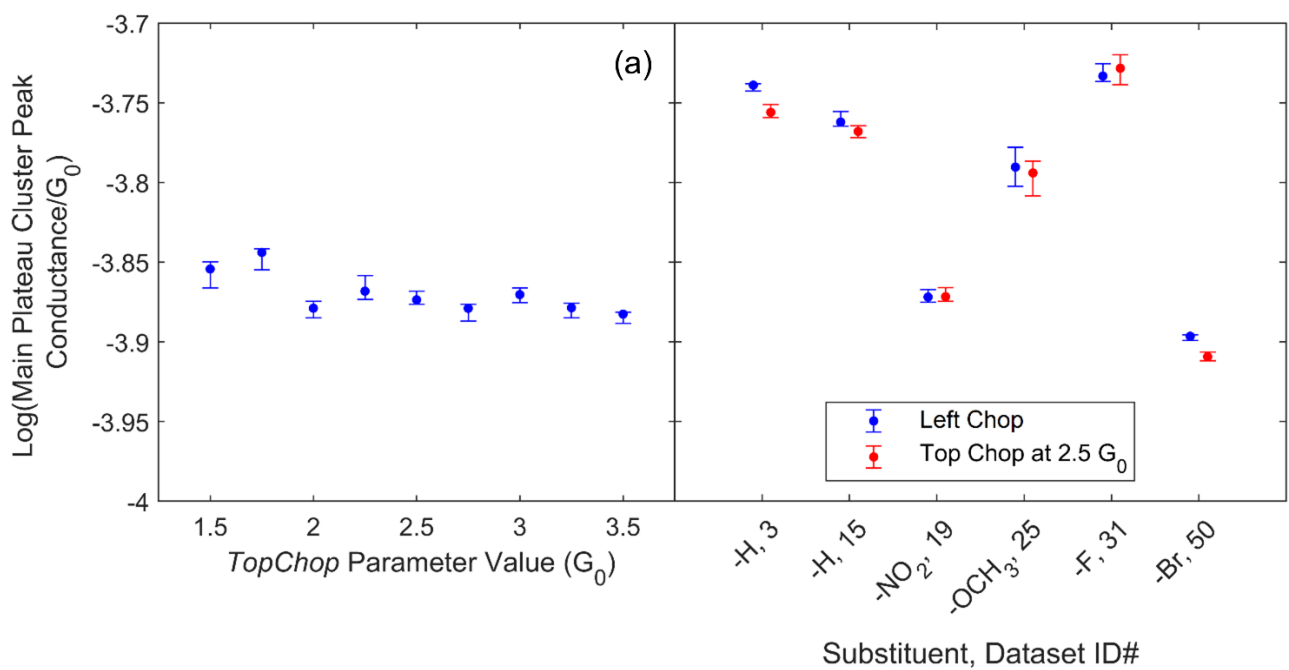

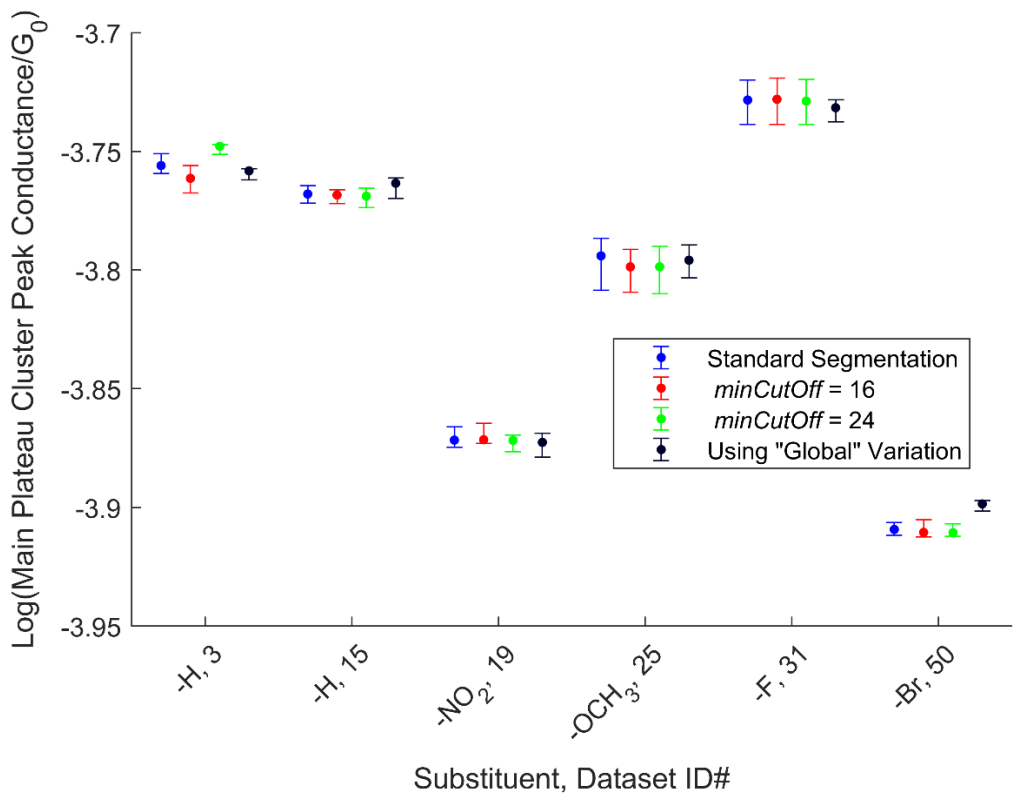

**Figure S5**, **Figure S7**, and **Figure S17**), we use the median from among these twelve peak values, along with error bars representing the range of the middle eight of the twelve values (i.e. the middle 66.7%).

*S.7 Selection of Main Plateau Clusters for OPV3-2BT-X Datasets*

Of the 43 OPV3-2BT-X datasets listed in **Table S2**, one dataset (ID# 114) did not produce any full-valley clusters that came close to corresponding to the molecular feature in the 2D histogram (possibly because the percentage of junctions containing a molecule was too low), and so was excluded from subsequent analysis. In 31 cases, only a single full-valley cluster had any similarity to the molecular feature, and each of these clusters was quite similar to the main plateau cluster shown in Figure 4h. We therefore unambiguously assigned each of these clusters as the analogous "main plateau cluster" for their respective datasets.

In 10 of the OPV3-2BT-X datasets, two full-valley clusters were found which might correspond well to the molecular feature region in the 2D histogram. However, in each of these cases, one of the clusters consisted of mostly flat segments like the main plateau clusters in the 31 datasets mentioned above (*e.g.* **Figure S8a,d,g,j**), whereas the second cluster consisted of more angled segments at slightly higher conductance (*e.g.* **Figure S8b,e,h,k**). Moreover, the valley corresponding to each flatter cluster always showed up in a similar location in its reachability plot as the other identified main plateau clusters (*e.g.* **Figure S8c,f,i,l**), suggesting that it represents an analogous component of the dataset's hierarchical structure. Therefore, in these 10 datasets there was still a single unambiguous choice for which full-valley cluster was the analogous feature to the cluster in Figure 4h and should thus be assigned as the main plateau cluster. **Figure S8** compares the chosen main plateau clusters with the angled clusters for four examples from these 10 datasets to demonstrate how clear these choices were.

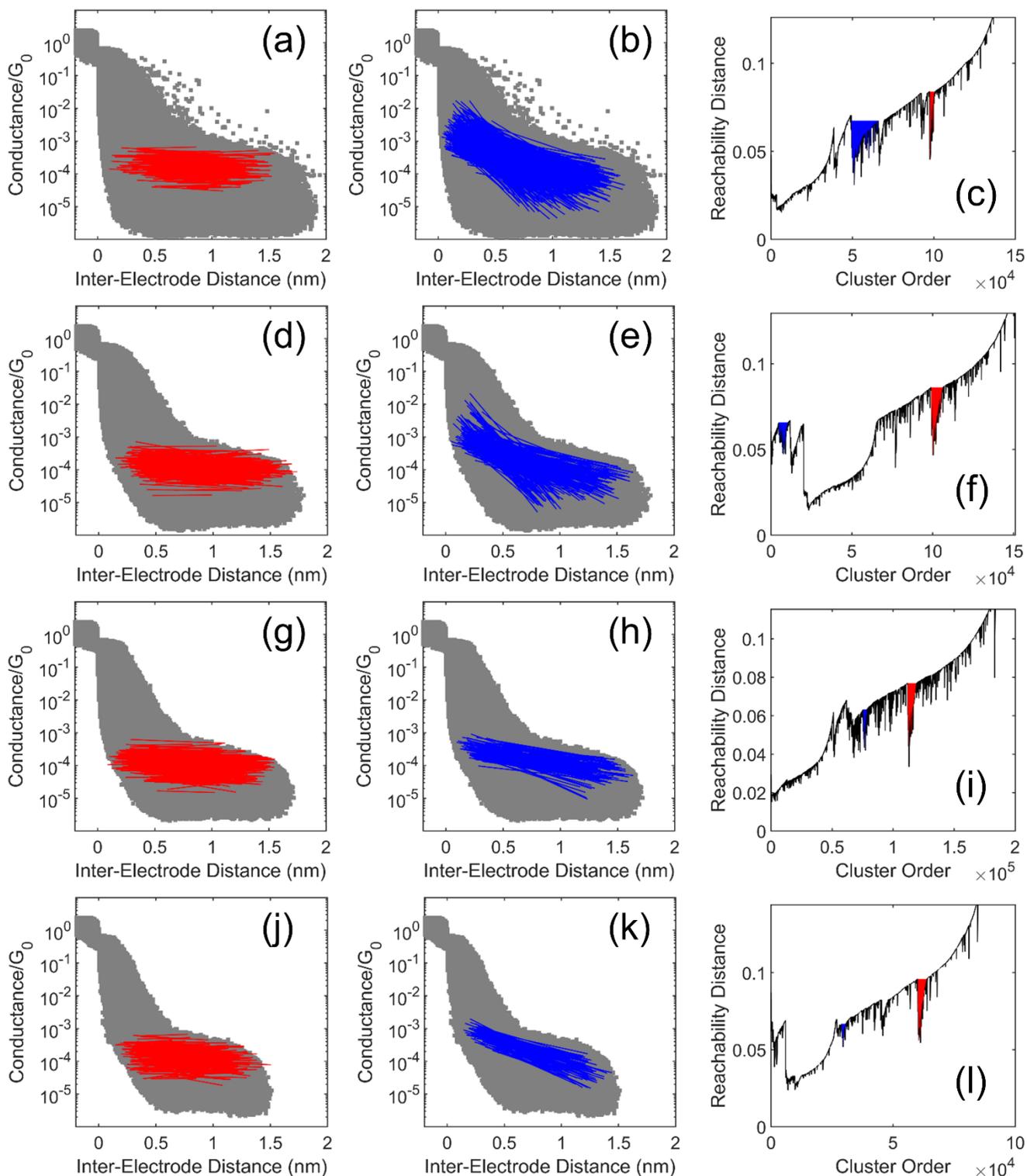

**Figure S8.** (a) Main plateau cluster chosen for dataset #59 (see **Table S2**). (b) Second full-valley cluster discovered in dataset #59 which corresponds well with the molecular feature from the 2D histogram, but is qualitatively distinct from the other identified main plateau clusters due to its higher conductance and more-angled segments. (c) Reachability plot for dataset #59 with the valleys corresponding to the clusters in (a) and (b) highlighted, showing how they fit into the hierarchical clustering structure. (d-f) Analogous plots for dataset #103. (g-i) Analogous plots for the dataset #104. (j-l) Analogous plots for dataset #37. Together, these four examples demonstrate that even in the datasets containing multiple molecule-like full-valley clusters, there was consistently an unambiguous choice for which cluster was structurally most analogous to the cluster in Figure 4h and should thus be assigned as the main plateau cluster (*i.e.*, the flatter clusters in the first column).

Finally, in one OPV3-2BT-H dataset (ID# 58), only a single full-valley cluster corresponding to the molecular feature was found (**Figure S9**), but this cluster resembled the angled clusters discussed above much more than the main plateau clusters identified in the other 41 datasets. This is therefore the second OPV3-2BT-X dataset that we excluded from subsequent analysis because this cluster does not appear to belong in the same category as the other 41. No qualitative change to our conclusions would have resulted from inclusion of this dataset.

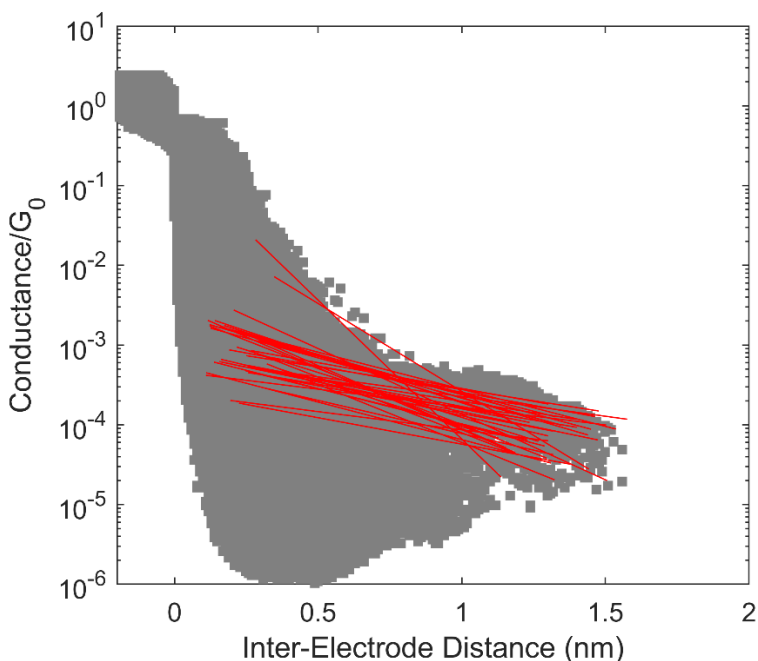

**Figure S9.** The only full-valley cluster from dataset #58 (see **Table S2**) which corresponds to the molecular feature in the 2D histogram. Because this feature seems to match the "secondary", angled clusters in **Figure S8** more than all other chosen main plateau clusters, it was excluded from subsequent analysis.

It is intriguing to note that the higher-conductance, more-angled clusters discovered in the 11 datasets discussed above appear qualitatively similar to the "class 2" traces identified by Cabosart et al. for a structurally similar molecule using a completely different clustering approach. In an additional similarity, Cabosart et al. also found a lower-conductance, flatter cluster ("class 3" traces) which they assign to the "standard" binding configuration and find to be a consistent representation of the molecular conductance.[10] This perhaps suggests that these two features might be a conserved motif of rod-like conjugated molecules, and full atomistic calculations are needed to investigate this question in more detail. On a more general level, the fact that significantly different clustering methods identify similar molecular features supports the view that clustering analysis is an appropriate means of revealing intrinsic data structure.

*S.8 Peak Fitting*

In order to have a point of comparison to our main plateau cluster peak fits, we pursued the standard approach of fitting the molecular peak in each raw 1D histogram with a single Gaussian. However, due to the complex and asymmetric peak shape, fitting within the conductance range surrounding the molecular peak typically leads to unreasonable results (*e.g.* dotted green line in **Figure S10**), and moreover can strongly depend on exactly how this conductance range is defined. Therefore, to fit the raw 1D histogram molecular peaks for our OPV3-2BT-H datasets, we used an iterative approach to set the conductance bounds for fitting. Each histogram is first fit with a single Gaussian peak while only considering the conductance range -5.5 $G_0$ to -2.5 $G_0$ (*e.g.* the dotted green line in **Figure S10**). Ten more restricted fits are then performed, with the conductance bounds modified each time based on the results of the previous fit. At each iteration, the conductance bounds are centered around the peak value from the previous fit, and the width of this fitting region is 2 decades for the first two iterations, 1.5 decades for the next four, and 1 decade for the last four. This process was empirically found to produce reasonable fits for the eight OPV3-2BT-H datasets we applied it to (*e.g.* dashed red line in **Figure S10**), and the peak value always fully converged by the tenth iteration.

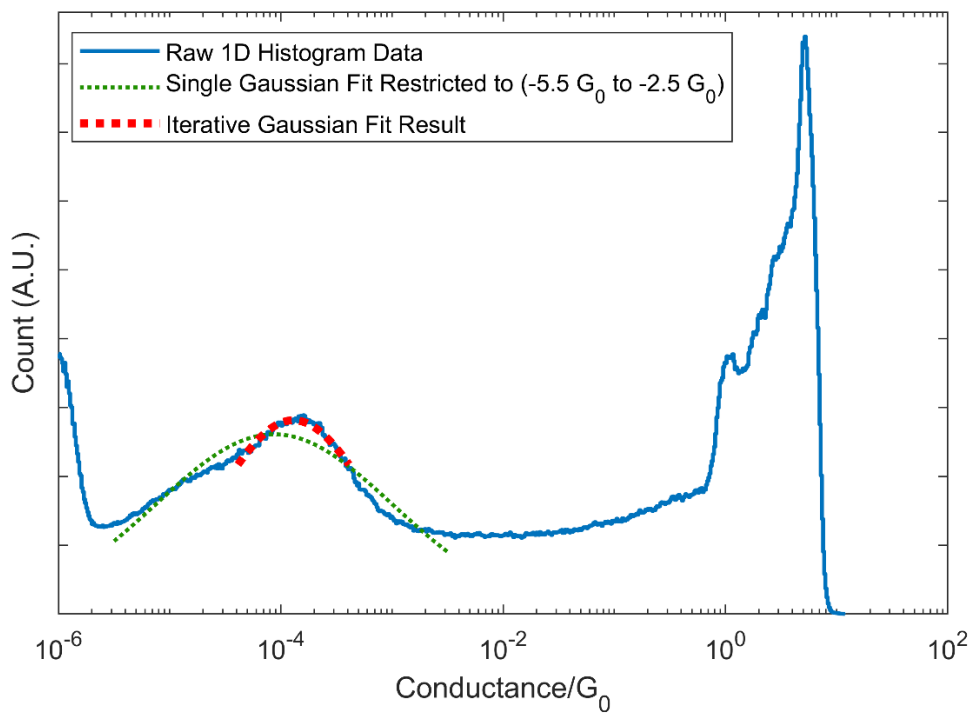

**Figure S10.** Raw 1D histogram for the OPV3-2BT-H dataset from Figure 1 (blue), along with a single Gaussian fit to only the range -5.5 $G_0$ to -2.5 $G_0$ (dotted green), and the result of an iterative process described in the text for determining the fitting range (dashed red).

For fitting the distributions of conductance values from specific clusters, in every case we used a single, unrestricted Gaussian fit. In the majority of cases, these distributions matched a Gaussian peak shape extremely well (e.g. Fig. 5). Some of the distributions displayed minor asymmetry or increased kurtosis, and thus fit a Gaussian peak shape less well; **Figure S11** shows the worst examples from the OPV3-2BT-X datasets. However, even in these cases, the single unrestricted Gaussian fit provided very reasonable approximations to the peaks and peak centers. A more complex fitting function would likely tighten the distributions of peak values in Figure 7; for example, adding a second fitting peak for the OPV3-2BT-Br and OPV3-2BT-Cl main plateau cluster distributions shown in **Figure S11c** and **Figure S11e**, respectively, would increase the conductance of the "main" peak, and these two datasets are both mild outliers on the low side in Figure 7.

For all histogram fitting in this work, the histogram bin width was determined based on the Freedman–Diaconis rule.

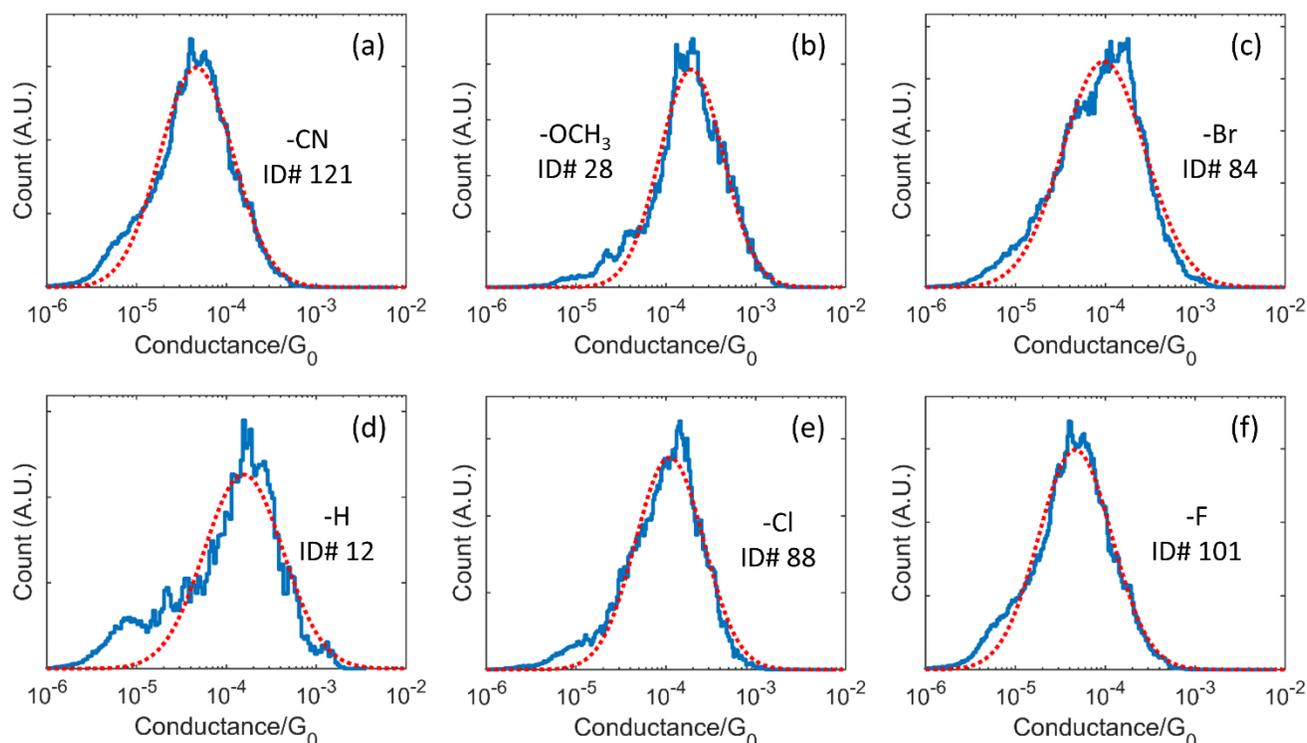

**Figure S11.** Main plateau cluster distributions (blue) and their respective unrestricted Gaussian fits (dotted red) for the six OPV3-2BT-X datasets in which these distributions were least Gaussian-shaped. The substituent, -X, and the ID# (from **Table S2**) for each dataset are inset for each plot.

*S.9 Investigating OPV3-2BT-X Main Plateau Cluster Lengths*

To help support our hypothesis that the main plateau cluster for each OPV3-2BT-X dataset represents the primary molecular feature, we investigated the maximum junction gap sizes implied by these clusters with two similar approaches. In the first method, we focus only on the actual trace pieces represented by the segments in the main plateau cluster. The end points of these trace pieces represent the maximum extent of each identified molecular plateau. However, it is possible that the linear features identified by Segment Clustering do not represent the entire time the molecule spent in the junction (e.g. the conductance may vary significantly during the detachment process). Therefore, in the second method we consider each entire trace containing a segment assigned to the main plateau cluster. The last time each trace drops to a low value well below the conductance of the molecule (here the value of $5·10^{-6}$ $G_0$ is used) is an alternative way to represent the distance at which the molecule fully breaks off. Both methods are demonstrated for an example OPV3-2BT-H dataset in **Figure S12a-d**.

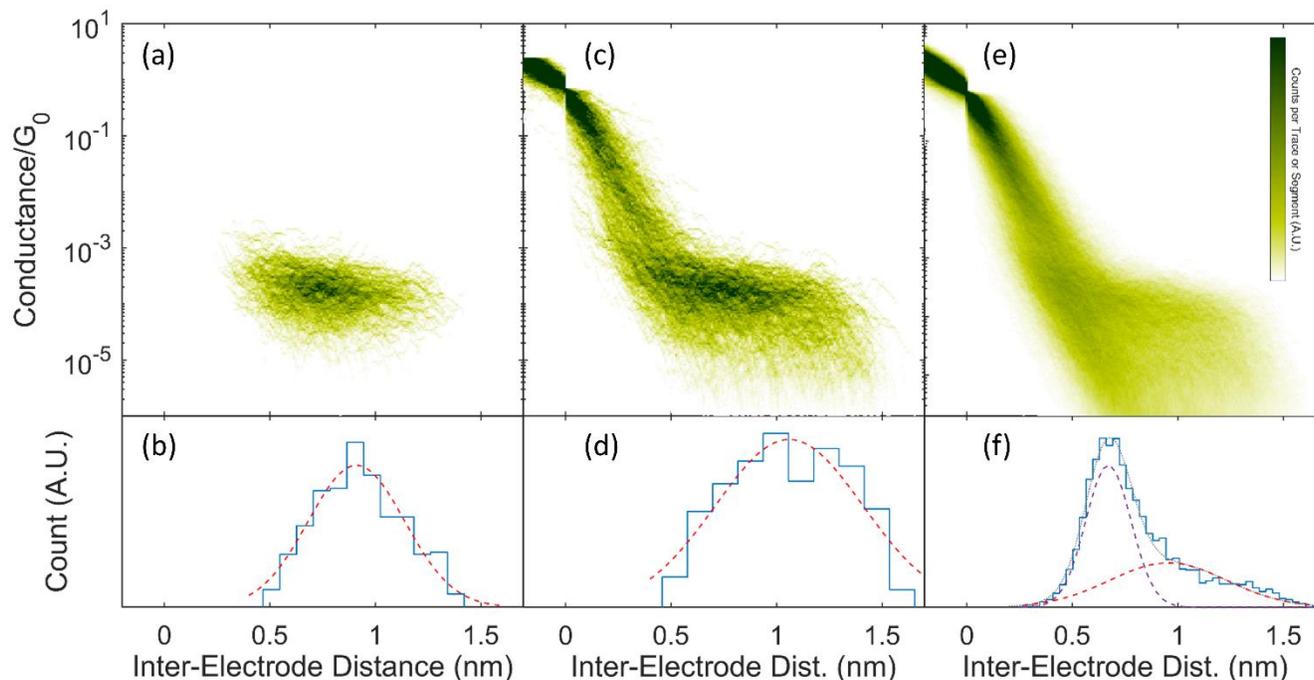

**Figure S12.** Examples of distance investigation methods using the OPV3-2BT-H dataset from Figure 1c. (a) 2D histogram of just the trace pieces whose linear segments were assigned to the main plateau cluster. (b) 1D histogram of the endpoints of the trace pieces in (a), fit with a single Gaussian peak (red). (c) 2D histogram of all traces containing segments which were assigned to the main plateau cluster. (d) 1D histogram of the distances at which each trace in (c) last crossed below the conductance value $5·10^{-6}$ $G_0$, fit with a single Gaussian peak (red). (e) 2D histogram of *all* traces in the dataset. (f) Analogous to (d), but for the traces shown in (e); fit with two Gaussians (purple and red, total fit in gray).

For comparison, we also show the results of applying the "trace-cross" method to *all* traces in the dataset (**Figure S12e,f**). This entire-dataset distance distribution exhibits two peaks, typically attributed to the break-off of tunneling traces and to molecular traces respectively.[11–14] As shown in **Table S5**, both distance distributions for the main plateau cluster are quite similar to the second peak in the entire-dataset distribution, providing clear evidence that what we label the "main plateau cluster" corresponds to what is generally considered to be the "primary" molecular feature. Similar results were obtained for the other OPV3-2BT-X datasets considered in this work. The moderate variation that was observed between

S31

datasets is likely due in large part to small systematic errors in attenuation ratios, and the overall pattern did not suggest any systematic differences in length between different substituents.

The fairly broad distributions seen in **Figure S12b,d** indicate that not all junctions reach the same degree of elongation before breaking off. The distribution peaks are somewhat shorter than what would be expected for fully-elongated molecular junctions, which is consistent with previous results for molecules with –BT linker groups.[15] This suggests that molecules with this linker group may in general not reach full extension.

**Table S5.** The peak and half-width at half-maximum (HWHM) values for the red Gaussian fits shown in **Figure S12** panels b, d, and f, respectively.

|  | Peak (nm) | HWHM (nm) |
|---|---|---|
| Segment End Points | 0.91 | 0.27 |
| Segment-Containing Trace Crosses | 1.06 | 0.41 |
| All Trace Crosses | 0.95* | 0.33* |

*For the higher-distance of the two Gaussian fits (red in **Figure S12f**).

*S.10 Selection of Main Plateau Clusters for OPV2-2BT and C6-2SMe*

**Figure S13** shows all of the full-valley clusters discovered in the OPV2-2BT dataset from Figure 8d. The cluster in **Figure S13i** can be unambiguously chosen as the main plateau cluster for the high-conductance feature. None of the full-valley clusters corresponds well to the low-conductance feature in this dataset (the cluster in **Figure S13f** is the closest, but does not align well with the low-conductance feature on either axis in the 2D histogram). Similar main plateau clusters were identified in the other four OPV2-2BT datasets considered in this work (**Table S3**).

S32

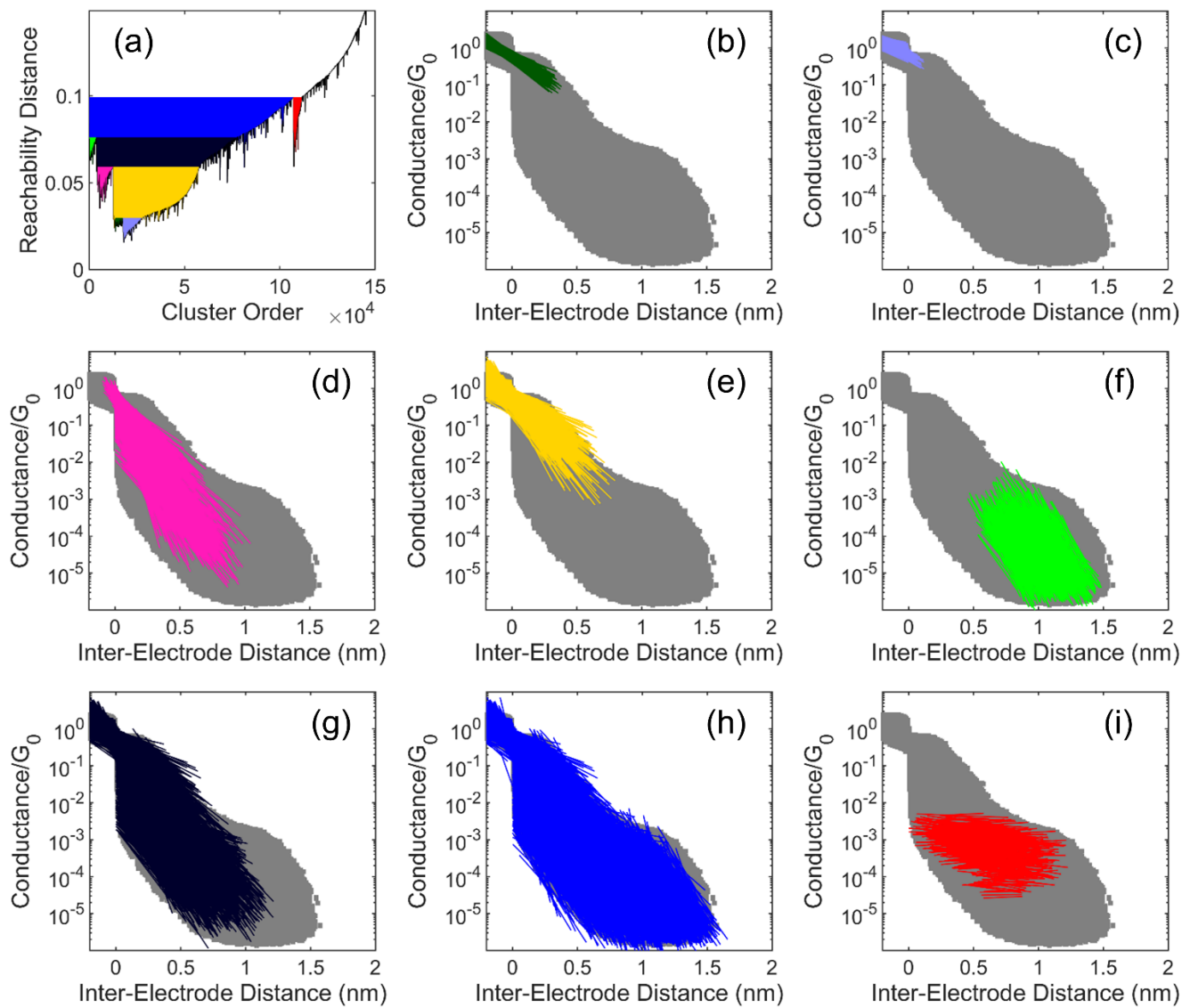

**Figure S13.** (a) Reachability plot for the OPV2-2BT dataset from Figure 8d with all full-valley clusters hierarchically filled in. (b-i) Segment clusters for each color coded valley from (a), with the cluster in (i) unambiguously identified as the main plateau cluster.

**Figure S14** shows full-valley clusters for the C6-2SMe dataset from Figure 8a. The cluster in **Figure S14l** can be unambiguously chosen as the main plateau cluster for this dataset. While the cluster in **Figure S14k** bears a superficial resemblance to the molecular feature, closer inspection reveals that it is much smaller and is composed of very angled segments which are unlikely to correspond to clean molecular plateaus. A similar main plateau cluster to **Figure S14l** was identified in the other C6-2SMe dataset considered in this work (**Table S3**).

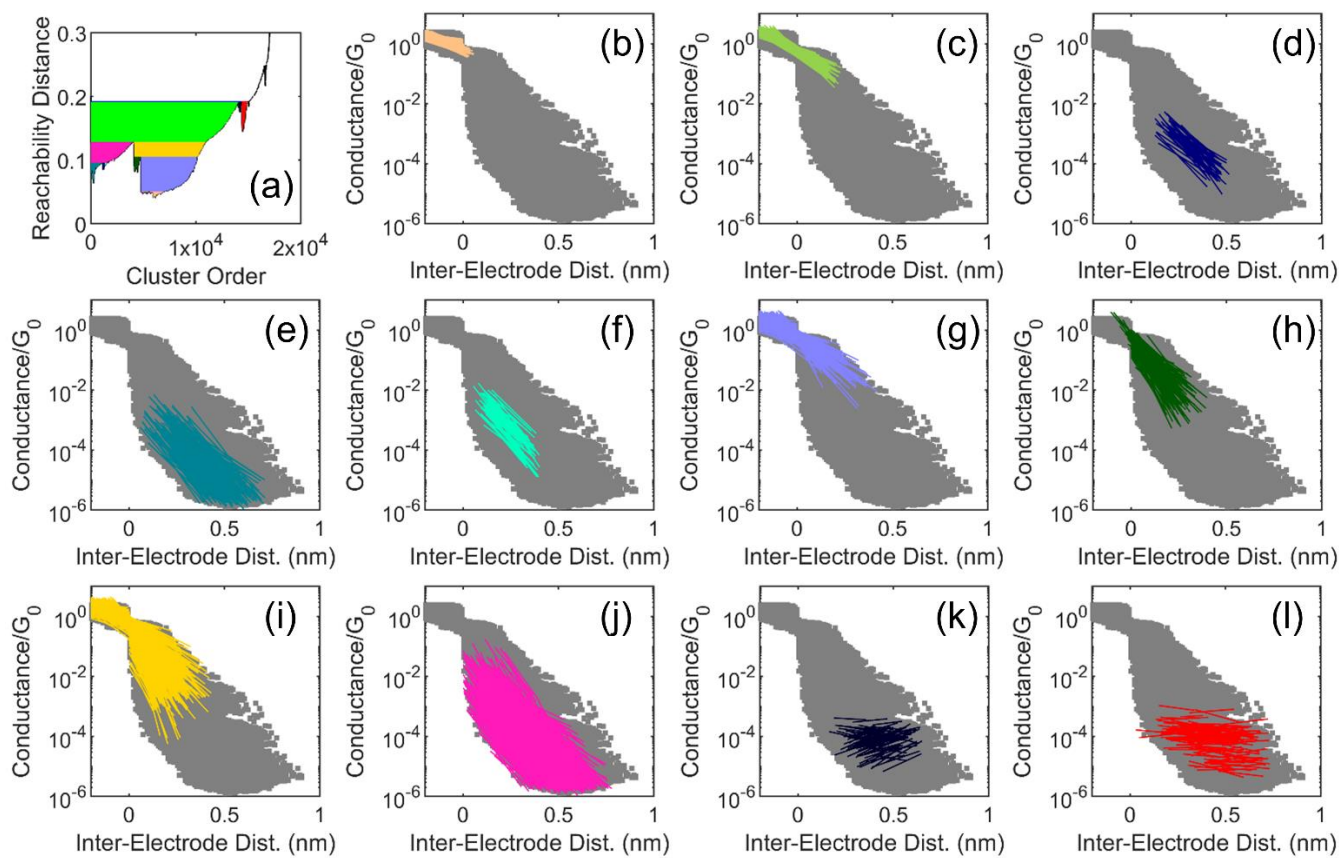

**Figure S14.** (a) Reachability plot for the C6-2SMe dataset from Figure 8a with all full-valley clusters hierarchically filled in. (b-l) Segment clusters for most of the color coded valleys from (a) (less-important clusters omitted for clarity), with the cluster in (l) identified as the main plateau cluster.

*S.11 Cluster Selection for OPV2-2BT/C6-2SMe 1:1 Synthetic Mixture #1*

When finding all full-valley clusters for a dataset, the minimum valley size should be set according to the specific context and what types of clusters the user is interested in. For the pure molecular datasets considered in this work, we found that a minimum valley size of 1% of the total number of data points worked well. However, in our synthetic mixture datasets each molecular feature is "diluted" by a factor of two. Moreover, because the C6-2SMe feature is so short, it represents a relatively small number of data points. Therefore, in this context a smaller minimum valley size is appropriate. To demonstrate this,

**Figure S15** shows full-valley clusters from the "Mixture #1" dataset from Figure 8g with a minimum valley size of 1%. Although a main plateau cluster can be easily identified (**Figure S15o**), this cluster contains features from both molecules. However, if the minimum valley size is lowered to ~0.5%, then the hierarchical structure produced by Segment Clustering reveals that the cluster from **Figure S15o** is composed of two main sub-valleys (**Figure S15p**). These two sub-valleys represent the clusters shown in Figure 8h, and, as discussed in the main text, correspond to the two different molecular features.

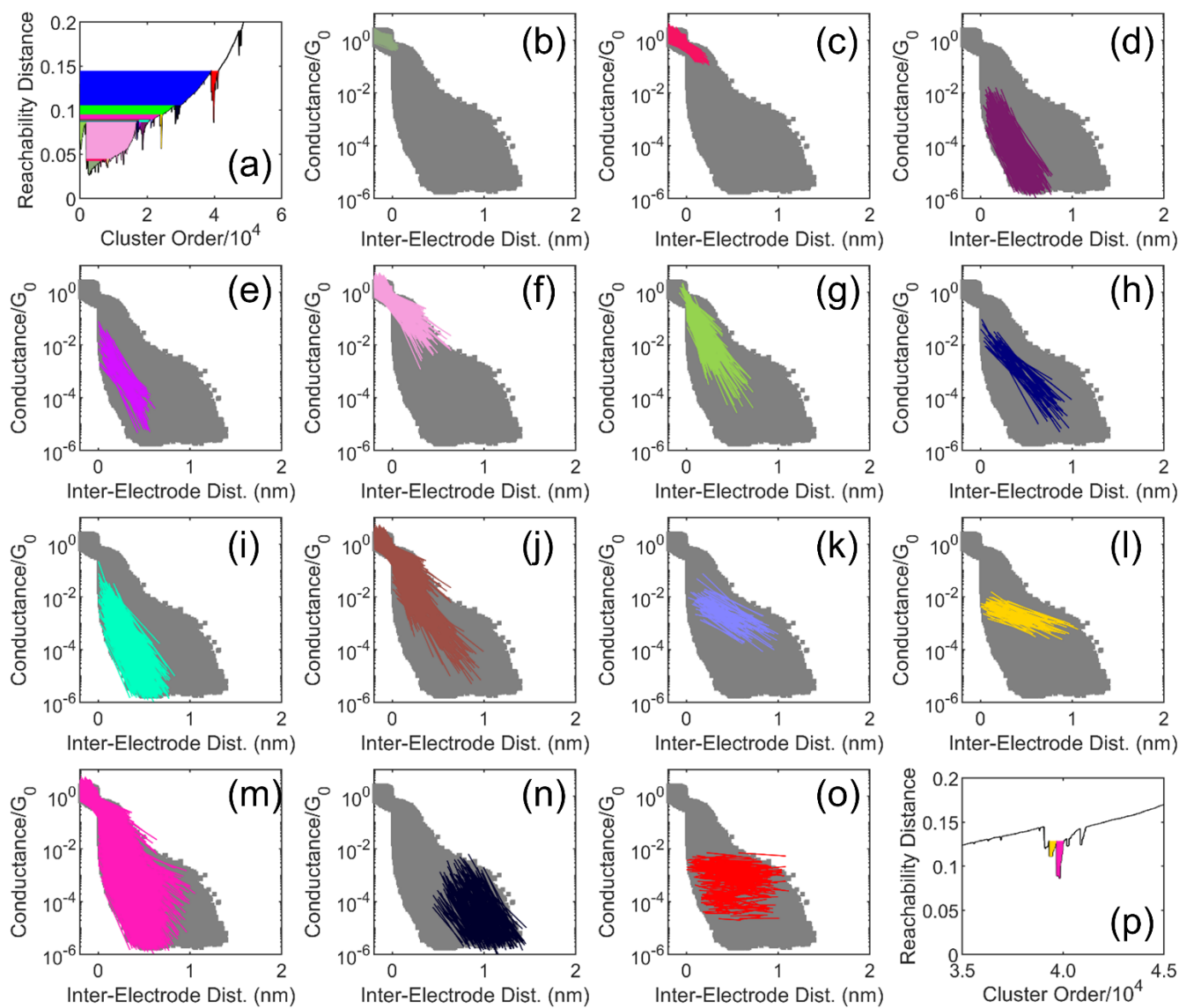

**Figure S15.** (a) Reachability plot for the Mixture #1 dataset (Fig. 8g) with all full-valley clusters hierarchically filled in. (b-o) Segment clusters for most of the color coded valleys from (a) (less-important clusters omitted for clarity). The red cluster in (o) is a composite plateau cluster for both molecular features. (p) By lowering the minimum valley size, the cluster in (o) is found to have substructure consisting of two separate valleys, corresponding to the two clusters plotted in Figure 8h.

*S.12 Clustering of Additional Synthetic Mixtures*

In addition to the OPV2-2BT/C6-2SMe mixture dataset discussed in the main text, 7 additional 1:1 synthetic mixture datasets (for a total of 8) were constructed (see **Table S4** for details) and analyzed in the same way. In seven of these eight total cases, two full-valley clusters were identified that correspond to the main OPV2-2BT and C6-2SMe molecular features (**Figure S16**). Just as with mixture #1 (see section S.11), in each of these cases a "composite" main plateau cluster was first unambiguously identified at the 1% valley size cut-off (analogous to **Figure S15o**), and then lowing of this cut-off revealed two primary sub-valleys (analogous to **Figure S15p**) corresponding to the two molecular features. The clusters identified in this way are shown in **Figure S16**, and their sizes are listed in **Table S6**. The one exception was Mixture #6, where the plateau cluster contained both molecular features did not possess any hierarchical sub-structure (**Figure S16f**). This illustrates the potential drawback of density-based clustering methods mentioned in the main text that dissimilar groups of data may in some cases end up in a single cluster if there is a continuous spread of data between them. We speculate that this issue occurs for this dataset because an error in the attenuation ratios results in similar apparent lengths for both molecules.

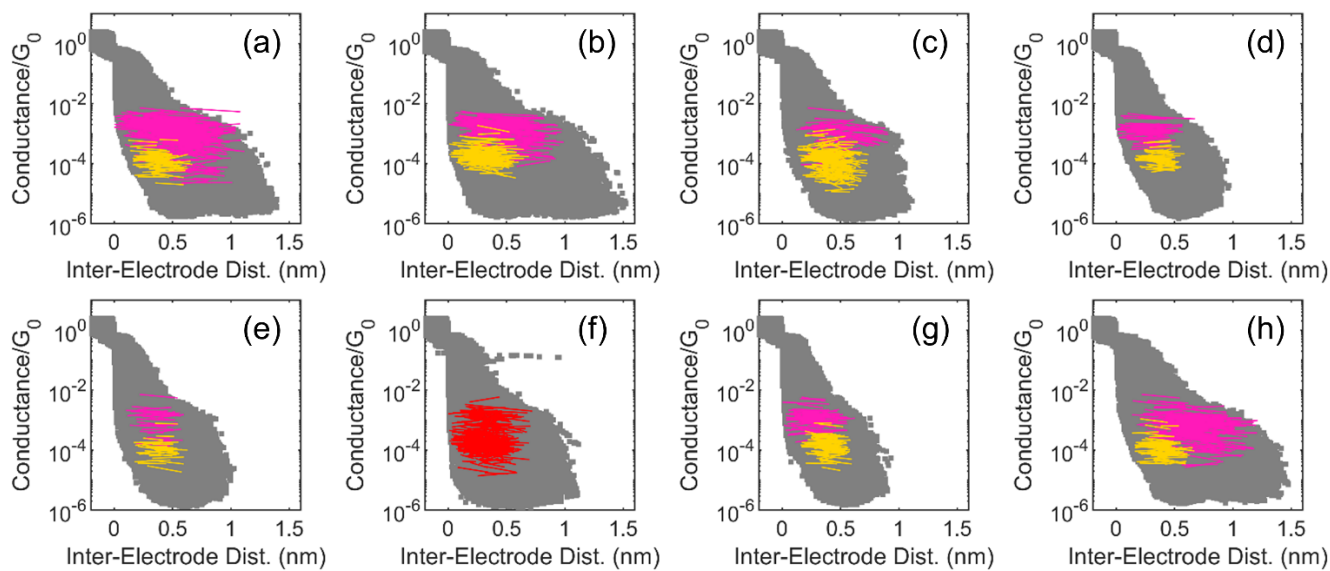

**Figure S16.** (a-h) Identified molecular plateau clusters for synthetic mixtures #1-8, respectively. In each case, a composite molecular plateau cluster analogous to **Figure S15o** was unambiguously identified (not shown). In 7 out of 8 cases, the valley for that composite cluster was found to contain two sub-valleys, analogous to **Figure S15p**, which were assigned as the OPV2-2BT plateau cluster (pink) and the C6-2SMe plateau cluster (yellow). As shown in **Figure S17** and **Table S6**, these assignments proved to be quite accurate, demonstrating the robustness of Segment Clustering's ability to separate overlapping molecular features. The composite cluster for mixture #6, shown in red in (f), did not contain any hierarchical sub-structure, and so could not be separated.

Just as with Mixture #1 in the main text, each of the OPV2-2BT (C6-2SMe) clusters in **Figure S16** was evaluated for accuracy by calculating how many of the data points assigned to it were from the traces belonging to the OPV2-2BT (C6-2SMe) half of the mixture (**Table S6**). This demonstrates that these separations of overlapping features were successful. While the C6-2SMe clusters again appear to display higher "error rates", as explained in the main text, this is unsurprising given the shorter plateaus for this molecule; the fact that a cluster of short C6-2SMe-like segments is not found in any of the pure OPV2-2BT datasets demonstrates that the source of the "erroneously" included segments is random chance, not mistaken feature identification by the algorithm. Finally, to summarize all of these mixture separation results, **Figure S17** compares the peak conductance values for the two identified molecular clusters from each mixture dataset with the peak conductance values from the main plateau clusters in the pure OPV2-2BT and C6-2SMe datasets.

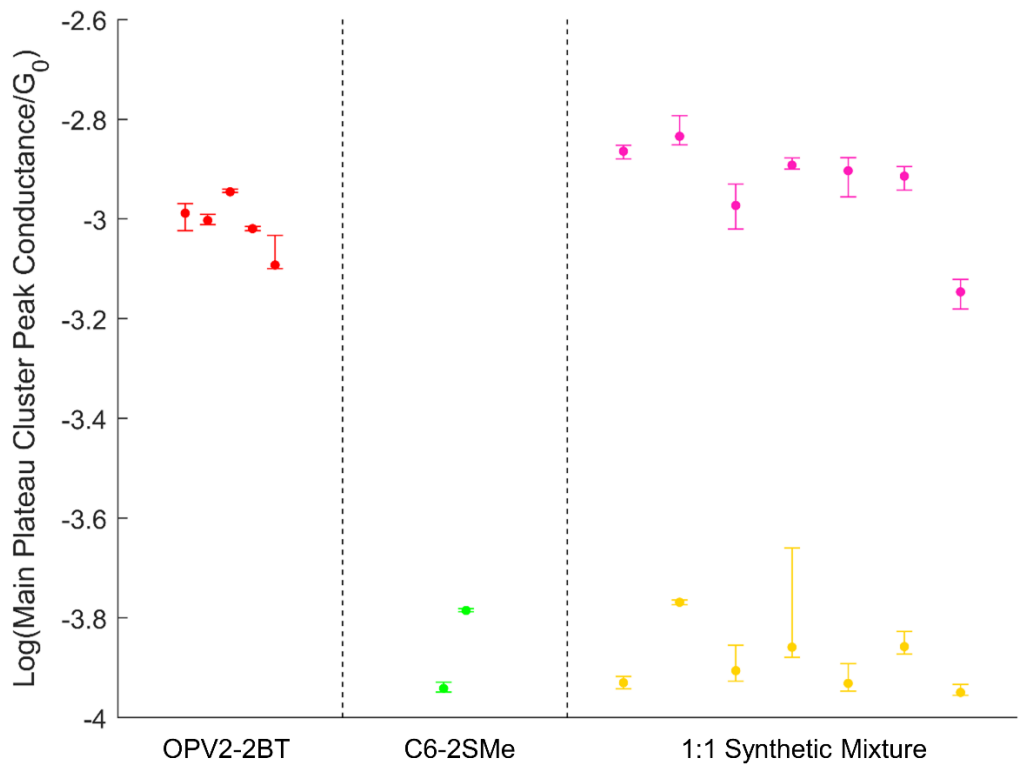

**Figure S17.** Peak conductance values for the main plateau clusters for the 5 different pure OPV2-2BT datasets considered in this work (red) and the two pure C6-2SMe datasets considered in this work (green). For comparison are plotted the peak conductances of the OPV2-2BT (pink) and C6-SMe (yellow) clusters identified in the seven successfully separated 1:1 synthetic mixture datasets shown in **Figure S16**.

**Table S6.** For each of the eight mixture datasets considered in this work, the size of the identified C6-2SMe and OPV2-2BT clusters (as a percentage of total data points) and the "accuracy" of each cluster (i.e. how many data points belonging to the cluster come from traces collected in the presence of the molecule that the cluster is assigned to). Each value represents the median from among the twelve different clustering outputs (using different values of the *minPts* parameter) for each dataset. Separate C6-2SMe and OPV2-2BT clusters could not be identified for mixture #6

| Mixture # | Data Points Contained in Cluster | | Data Points from Correct Half of Dataset | |
|---|---|---|---|---|
| | C6-2SMe | OPV2-2BT | C6-2SMe | OPV2-2BT |
| 1 | 0.5% | 3.2% | 84% | 97% |
| 2 | 1.4% | 1.6% | 84% | 99% |
| 3 | 1.4% | 0.3% | 60% | 91% |
| 4 | 0.4% | 0.9% | 67% | 98% |
| 5 | 1.0% | 0.4% | 69% | 90% |
| 6 | NA | NA | NA | NA |
| 7 | 0.8% | 0.7% | 59% | 96% |
| 8 | 0.5% | 1.8% | 76% | 95% |
| AVERAGE | 0.9% | 1.3% | 71% | 95% |